\documentclass[journal=jacsat,manuscript=article]{achemso}

\usepackage{xcolor}
\usepackage{graphicx} 
\usepackage{dcolumn} 
\usepackage{bm} 
\usepackage{hyperref}
\hypersetup{
    colorlinks=true,
    linkcolor=red,  
    citecolor=black,
    urlcolor=cyan,
    pdftitle={Overleaf Example},
    pdfpagemode=FullScreen,
    }
\usepackage{enumitem}
\usepackage{soul}
\usepackage{amsmath}
\title{Ionic Liquid-Driven Modulation of DNA Brush Morphology on Nanoparticle Surfaces}%

\author{Anuj Chhabra}
\altaffiliation{These authors contributed equally to this work.}
\affiliation[Unknown University]
{Center for Research in Nanotechnology $\&$ Science, Indian Institute of Technology Bombay, India}
\author{Sandip Mandal}
\altaffiliation{These authors contributed equally to this work.}
\affiliation[Unknown University]
{Center for Condensed Matter Theory,
Department of Physics, Indian Institute of Science, Bangalore 560012, India}
\author{Yugang Zhang}
\affiliation[Unknown University]
{Center for Functional Nanomaterials,
Brookhaven National Laboratory, Upton, New York}
\author{Oleg Gang\textsuperscript{*}}
\affiliation[Unknown University]
{Center for Functional Nanomaterials,
Brookhaven National Laboratory, Upton, New York}
\alsoaffiliation[Unknown University]
{Department of Chemical Engineering, Columbia University, New York}
\alsoaffiliation[Unknown University]
{Department of Applied Physics and Applied Mathematics,Columbia University, New York}
\author{Prabal K. Maiti\textsuperscript{*}}
\affiliation[Unknown University]
{Center for Condensed Matter Theory,
Department of Physics, Indian Institute of Science, Bangalore 560012, India}
\author{Sunita Srivastava}
\affiliation[Unknown University]
{Soft Matter and Nanomaterials Laboratory, Department of Physics, Indian Institute of Technology Bombay, Powai, Mumbai 400076, India}
\email{sunita.srivastava@iitb.ac.in;maiti@iisc.ac.in;ogang@bnl.gov}
\phone{+91 (0) 022 2576 7572}

\date{\today}

\begin{document}

\begin{abstract}
The morphology of DNA is strongly influenced by its surrounding environment, including factors such as pH, salt type and valency, and the presence of polymers. Inorganic salts are known to reduce the DNA chain length through mechanisms like electrostatic screening and ion bridging. In contrast, ionic liquids, a new class of organic salts, have previously been found to increase the DNA chain length, indicating a distinct mode of interaction between the ionic liquid and DNA chains. This study utilizes self-assembled DNA-AuNPs as a model system to examine changes in the DNA chain morphology and the nanoscale interaction mechanisms in ionic liquid environment. The DNA chain lengths are measured in solution using X-ray scattering measurements at varying concentrations of two imidazolium ([$BMIM$] acetate and [$EMIM$] acetate) based ionic liquids. Additionally, Molecular Dynamics (MD) simulations are performed mimicking the experimental system. Our results suggest an interplay of electrostatic and groove-binding interactions governing the DNA chain morphology, which depends on IL concentration and the composition of the DNA chains. It has been found that for DNA chains with majority ssDNA, electrostatic interaction dominate, however with increasing composition of double strands, the DNA chains exhibits compaction due to non-electrostatic hydrophobic groove-binding mechanism.
\end{abstract}
\maketitle
\section{\label{sec:level1}Introduction}
DNA, a negatively charged polyelectrolyte, are highly sensitive to environmental changes such as pH, salt valency, polymers etc\cite{rudiuk2012enhancement,srivastava2018liquid,meka2017comprehensive,innes2021flexible}. DNA exhibits a decrease in contour length with increasing salt concentration. In the case of inorganic salts, counterions interact with the DNA phosphate backbone, leading to reduced electrostatic repulsion between monomers and promoting attractive interactions that decrease the DNA persistence length\cite{luan2008dna, wang1991small,cruz2022twisting,aggarwal2020we}. As the concentration of monovalent salts increases, the counter-ions reduce the chain length by screening electrostatic repulsion \cite{srivastava2022effect}.
In case of multivalent counterions, other effects such as ion-bridging and overcharging come into play, which can lead to faster compaction or partial expansion of DNA chains due to charge inversion\cite{ghosh2015ion,wang2013salt, sreedhara2002structural}.\\ 
Beyond traditional salts, a new class of organic solvent known as Ionic Liquid (IL) is being studied. The interaction between IL and DNA has gained significant attention in recent years due to unique physiochemical properties of IL and there potential to provide DNA stability\cite{vijayaraghavan2010long}.
Ionic Liquids are salts that exist in a liquid state at room temperature \cite{zhao2015dna,egorova2021ionic}. It is often referred to as a green solvent due to its various properties, such as bio-compatibility, thermal stability, and low toxicity\cite{ghandi2014review}. 
Ionic liquids have demonstrated the potential for long term stability of dsDNA chains at room temperature\cite{vijayaraghavan2010long}, eliminating the need to store DNA below freezing point. Ionic liquids have also been shown to enhance the PCR amplification of GC-rich DNA, which is typically difficult to amplify\cite{shi2012ionic}. Additionally, they can help form DNA ionogels at DNA concentrations as low as 1 wt $\%$ \cite{pandey2017dna,pandey2018imidazolium}. Interestingly work by Maiti et. el. indicated the increase of the DNA persistence length with increasing IL ionic strength\cite{garai2018ionic}. This effect contrasts with what is observed for inorganic salts, making the interaction mechanism of ILs with the DNA chains a topic of interest. Chandran et. al. employed molecular dynamic simulations to demonstrate the interaction mechanism of ILs and the DNA chains. They observed that IL interacts both with the DNA backbone and with the DNA grooves.\cite{chandran2012groove}. This combination is absent for the inorganic salt and DNA interactions, which is majorly driven by the electrostatic interaction. Similarly, Jumbri et. al. corroborated these findings, showing that IL molecules can penetrate the DNA grooves and disrupt the DNA hydration shell\cite{jumbri2014insight}. IL molecules are found to disrupt the hydration shell around DNA chains, thereby providing enhanced stability to the DNA\cite{fadaei2022structural}. Also, fluorescence quenching experiments have further revealed that ILs can interact with the DNA grooves, replacing the fluorescence probes from the DNA hydration shell\cite{pabbathi2015spectroscopic,liu2016evaluation,ding2010binding}. Work by Fadaei et al. used a combination of fluorescence binding studies and molecular dynamics simulations to provide evidence for groove binding of imidazolium based ILs with DNA chains\cite{fadaei2021interactions}. Previous work has shown that additional interactions other than electrostatic interactions such as hydrophobic interactions need to be considered for IL interactions with the DNA chains\cite{jumbri2016binding,jumbri2019fluorescence,menhaj2012exploring,olave2024dna}. Although previous studies have explored DNA-IL interactions, investigations examining the morphological changes in DNA chains in the presence of ILs remain sparse. In this work, DNA functionalized nanoparticle self-assembly have been employed to study how IL interactions with DNA chains influence DNA chain morphology.\\
DNA has been established as a building block for creating complex nanostructures through programmed assembly \cite{ storhoff1999programmed,dey2021dna,li2025crystalline,nykypanchuk2008dna,xiong2009phase,macfarlane2020nanoparticle,jones2020programmable,mandal2014dna,kumar2012structure}. This programmability arises from the specific and predictable pairing of single-stranded DNA (ssDNA) to form double-stranded DNA (dsDNA) via Watson-Crick (WC) base pairing. DNA-gold nanoparticles (DNA-AuNPs) offer precise control over key parameters, including DNA chain length \cite{nykypanchuk2008dna}, particle size \cite{macfarlane2010establishing}, and the resulting lattice structures \cite{jones2010dna,srivastava2013super}. Previous research has explored the influence of environmental factors such as pH \cite{srivastava2018liquid} and salt concentration \cite{tian2016lattice} on the stability and assembly of DNA-AuNPs. These systems have been utilized to investigate the behavior of charged polymers in various salt environments, both in solution\cite{srivastava2022effect} and at interfaces \cite{luoAcsnano2011,SrivastavaAcsnano2014}.
Building upon the previous work on DNA nanoparticle self-assembly, our work aims to investigate how ionic liquids (ILs) affect the morphology and interactions of DNA within DNA-AuNP assemblies. By systematically varying the base composition (ss vs ds bases) of the DNA chains, we explore the interplay of electrostatics and groove binding interactions in presence of ILs. We employ a combination of Small Angle X-ray Scattering (SAXS) and Molecular Dynamics (MD) simulations to characterize these nanoscale interactions between the IL and the DNA chains.

\begin{figure*}
\centering
\setlength{\fboxrule}{1pt}
\includegraphics[width={1\textwidth}]{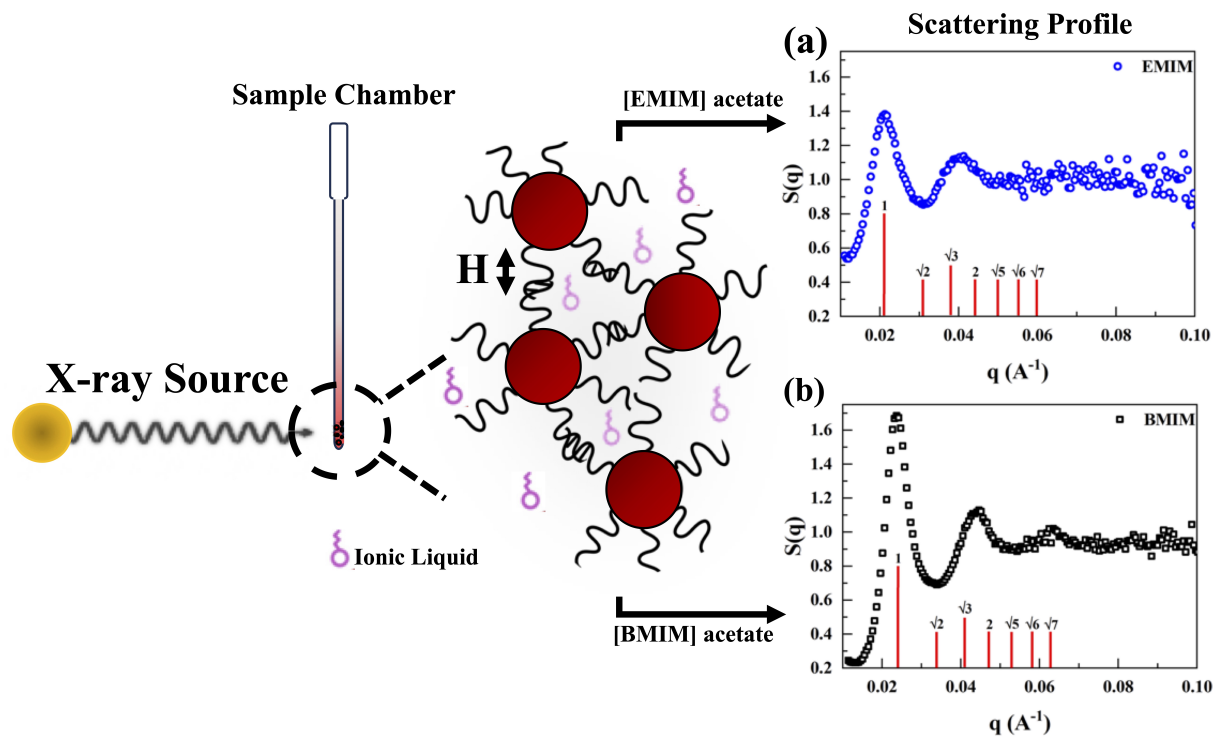}
\caption{ \textit{Schematic illustration of the Small Angle X-ray Scattering setup used to study the self-assembly structures of DNA-grafted gold nanoparticles in Ionic Liquid (IL) environments. The self-assembly behaviour is investigated in the presence of [$EMIM$] acetate and [$BMIM$] acetate. The corresponding structure factor, $S(q)$, as a function of scattering vector, $q$, are shown for a) [$EMIM$] acetate and b) [$BMIM$] acetate. The observed $S(q)$ indicates the formation of randomly ordered self-assembly.}}
\label{fig1}
\end{figure*}
\section{Results and Discussions}
\begin{figure*}
\centering
\includegraphics[width={1\textwidth}]{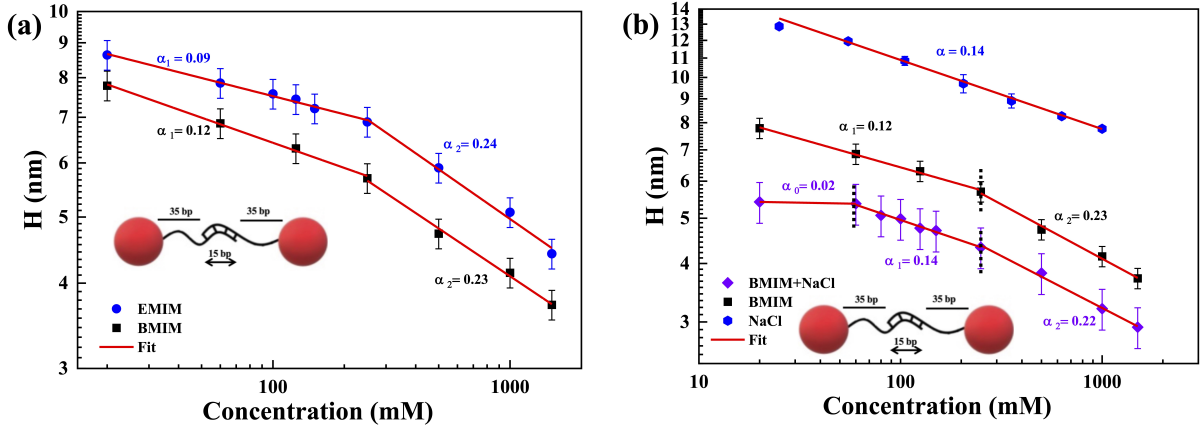}
\caption{ \textit{Dependence of DNA brush length, $H$ on the concentration of Ionic Liquid. a) Comparison of $H$ with increasing concentration of [$EMIM$] acetate and [$BMIM$] acetate. The experimental data is analysed using the power law ($H~\sim~C_{s}^{-\alpha}$) The fitted data shows different regimes with scaling exponent ($\alpha$). b) Effect [$BMIM$] acetate with added $NaCl$ compared to [$BMIM$] acetate and $NaCl$ alone. Insets illustrate the DNA-AuNP system corresponding to \textit{ds17}.}}
\label{fig2}
\end{figure*}

To measure the changes in DNA chain morphology Small Angle X-ray Scattering experiments were performed on DNA-gold nanoparticle (DNA-AuNP) self-assemblies in the presence of two types of ionic liquids at varying concentrations. A schematic of experimental setup is shown in Figure \ref{fig1}, where an X-ray source is directed on the self-assembled DNA-gold nanoparticles structure to obtain the scattering profile. The detailed experimental procedure is described in the Materials and Methods section. The structure factor ($S(q)$) profiles of SAXS data from assemblies of \textit{ds 17} (comprising gold nanoparticles functionalised with 50 bases single stranded DNA, of which 15 outer can form the Watson Crick base pairing to form self-assembly) in [$BMIM$] acetate and [$EMIM$] acetate, are shown in Figure \ref{fig1}a,b respectively. The observation of peaks in $S(q)$ indicates the formation of self-assembled structures. For DNA-AuNPs in [$BMIM$] acetate and [$EMIM$] acetate (Figure~\ref{fig1}), $S(q)$ shows broad peaks. Compared to the standard BCC lattice~\cite{jones2010dna}, the absence of sharp peaks suggests a random close-packed arrangement of DNA-AuNPs in the ionic liquid environment. For [$BMIM$] acetate, the first peak occurs at a higher \textit{q} compared to [$EMIM$] acetate at the same concentration, indicating reduced interparticle distances in the presence of [$BMIM$] acetate. The calculated structure factor for ds 17 at increasing IL concentration (Figure \ref{S3}) show the that with increasing both [$EMIM$] acetate and [$EMIM$] acetate concentration, the peak in \textit{S(q)} shifts to higher \textit{q} values, indicating a decrease in the interparticle distances of the DNA-AuNP self-assembled structures.\\
To estimate the length of DNA chains attached to the nanoparticle surface, we define the parameter \textit{H}, using the formula $H=\frac{d-2r}{2}$, where \textit{d} is the inter-planar spacing given by the equation $d = 2\pi$/q, \textit{q} is the position of the first peak in \textit{S(q) vs q} plot and $2r \sim 15~nm$ is the nanoparticle size. Importantly, the scattering profiles remained largely unchanged regardless of whether the assemblies were formed in the presence of ionic liquids alone (without added inorganic salts) or exclusively with inorganic salts (as shown in Figure \ref{S4} in the Supporting Information). In both cases, the data indicate the formation of a random closed packed assembly of DNA-AuNPs in ionic liquid environment.\\
Figure \ref{fig2} a presents the estimated values of $H$ at different concentrations of ionic liquid (IL). To ensure reproducibility, the measurements of $H$ were repeated at least three times, and the corresponding standard deviation in the estimates of H is shown as error bars. Two key observations emerge from this data: firstly, system with [$EMIM$] acetate exhibit consistently higher $H$ values compared to systems with [$BMIM$] acetate; and secondly, a non-monotonic decrease in \textit{H} is observed with increasing IL concentration for both systems, characterized by a change in slope. Previous studies on free DNA chains interacting with ILs, have indicated that decrease in DNA chain lengths are primarily governed by electrostatic screening and hydrophobic interactions (groove binding)\cite{jumbri2016binding,andrade2022imidazolium}. The positive charge on the benzene ring in the IL facilitates electrostatic screening, while the alkane chain contributes to groove binding through hydrophobic interactions. Since the positive cation for both [$BMIM$] acetate and [$EMIM$] acetate ILs are similar, the higher estimate of $H$ for [$EMIM$] acetate is attributed to it's weaker interaction with the DNA chains. Owing to the shorter alkyl chain of [$EMIM$], the reduced groove binding interactions can lead to an increase in $H$.\\
The power law analysis for the $H~vs~C_s$ data are performed by fitting the data to the equation, $H~\sim C_s^{-\alpha}$, where $\alpha$ is the power law exponent. Figure \ref{fig2} a shows the fitted data (solid line) and the corresponding estimates of the power law exponents. Notably, the analysis reveals distinct regimes of two different slopes, with a crossover occurring at approximately 250 $mM$. We termed the slopes before crossover as $\alpha_1$, and the slope after the crossover as $\alpha_2$. The estimated values of $\alpha_1$ and $\alpha_2$ for [$BMIM$] acetate are $0.12~\pm~0.01$ and $0.23~\pm~0.01$, respectively, while for [$EMIM$] acetate they are $0.09~\pm~0.01$ and $0.24~\pm~0.02$. Notably, the $\alpha_1$ values closely match those for a monovalent salt system containing \textit{NaCl} (Figure~\ref{fig2}a). Previous studies showed that electrostatic screening by metallic monovalent salts yields a power-law slope of approximately \textit{0.14} for DNA chains~\cite{srivastava2022effect}. The steeper slope $\alpha_2$ $\sim 0.24$ observed at higher concentrations (Figure~\ref{fig2}a) indicates a rapid decrease in $H$, suggesting the onset of stronger interactions. In contrast, prior work on divalent salts such as $MgCl_2$ exhibited high exponents due to ion bridging but with a single slope~\cite{srivastava2022effect}, differing from the two-slope behavior seen in ionic liquid systems. Whether a similar mechanism operates in ionic liquids at higher concentrations remains uncertain and is the subject of ongoing investigation to clarify the origins and interaction mechanisms behind this phenomenon.\\
\begin{figure*}
\centering
\includegraphics[width={1\textwidth},keepaspectratio]{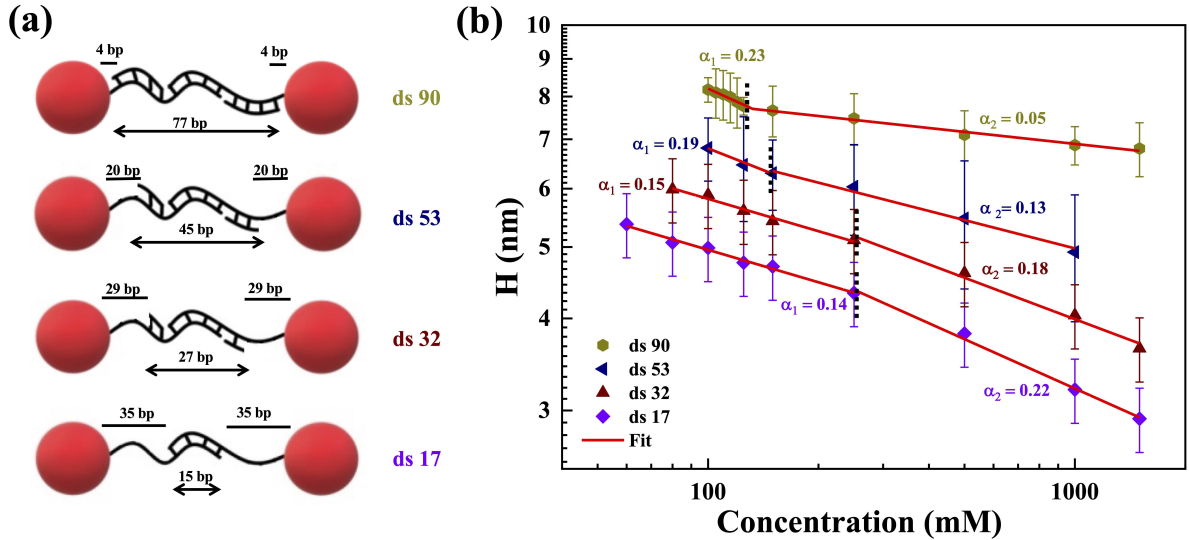}
\caption{\textit{ (a) Schematic representation of system design indicating the ratio of the dsDNA to ssDNA. The \textit{ds 17}, \textit{ds 32}, \textit{ds 53}, and \textit{ds 90} have dsDNA to ssDNA ratio of 15:70, 27:58, 45:40, and 77:8, respectively. Here \textit{ds 90} indicates presence of 90\% paired dsDNA nuclotides. (b) Dependence of length of DNA chain, $H$ on the concentration of [$BMIM$] acetate, analyzed by varying the dsDNA to ssDNA ratio.}}
\label{fig3}
\end{figure*}
In addition to electrostatic interactions, mechanism such as ion-bridging, and groove binding, intercalation  have also been reported to contribute to DNA compaction\cite{jumbri2019fluorescence,garai2018ionic,mandal2025vs}. In the presence of ionic liquids (ILs), these interactions have lead to the formation of complex DNA-IL structures, such as coiled structures or globules  \cite{gu2022ionic}. The structure of these complexes can vary depending on the specific properties of the IL and its interactions with DNA bases and the backbone. Electrostatic and groove binding interactions are the primary driving forces behind the observed reduction in DNA chain length when interacting with imidazolium based ILs\cite{jumbri2016binding}. The crossover of exponents depicted in Figure \ref{fig2}a provides evidence that the balance between electrostatic interactions and groove binding interactions leads to distinct scaling behaviour, with different slopes emerging in each regime depending on IL concentration.\\
To comprehensively investigate the interplay of these different interaction mechanisms, we have designed a series of systems with varying compositions of double-stranded nucleotides in the DNA chains (dsDNA). Our hypothesis proposes that through systematic adjustments to the double-stranded composition, we can modulate the competitive influences of electrostatic and hydrophobic interactions, thereby eliciting distinct power law exponents in the $H~vs~C_s$ dataset. Our system design requires the addition of a DNA linker to modify the dsDNA composition, and therefore, the presence of monovalent salt is necessary to form Watson Crick base pairing for linkers. Thus, prior to conducting the proposed experiments, it is imperative to verify whether the presence of monovalent salt (essential for creating double-stranded DNA strands) alters the response of DNA chain length, in the presence of [$BMIM$]. To address this, we performed experiments with added [$BMIM$] at a fixed concentration of $100~mM$ $NaCl$. Though it is known that the concentration of $100~mM$ $NaCl$ is sufficient for stable Watson-Crick interaction and formation of ds DNA chains\cite{tan2006nucleic}. We have verified through melting temperature estimates that the \textit{ds} connections are stable at room temperature and in $100 mM$ $NaCl$. 
For simplicity, in the \textit{ds17} system, where the dsDNA content is low, the interactions can be reasonably assumed to occur exclusively with ssDNA.
We compare the $H~vs~C_s$ graph for \textit{ds 17}, in the presence and absence of $100~mM$ $NaCl$ while varying the [$BMIM$] concentration. 
Two key observations emerge from the data in the presence of $NaCl$ (Figure \ref{fig2}b). (i) The onset of change in \textit{H}, occurs at a slightly higher concentration of [$BMIM$] acetate $\approx 60~mM$. (ii) Beyond $60~mM$ [$BMIM$] acetate, two distinct slopes are observed with power law exponents \textit{0.14} and \textit{0.22}, similar to the system without $NaCl$.
Qualitatively, we observed similar effects in $H~vs~C_s$ data for samples with and without $NaCl$. Previous studies suggest that if more than one type of counter-ion is present in the system, competition to bind to the DNA chain can occur \cite{yoo2012competitive}. This competition necessitates a higher threshold [$BMIM$] acetate concentration to effectively displace \textit{$Na^+$} from the DNA backbone. Consequently, the observed change in slope occurs only after reaching this critical concentration. Here, we will focus on the regions with two different slopes to understand interactions leading to such behavior in DNA chain morphology in the presence of ILs. \\
Figure \ref{fig3}a illustrates the design of distinct DNA systems, each containing varying percentages of double-stranded (ds) nucleotides ranging from $17~\%$ to $90~\%$ (Materials and Methods). Figure \ref{fig3}b presents the data exclusively within the range where [$BMIM$] acetate influences DNA chain length and where measurable changes in \textit{H} occur. The complete dataset for the entire salt concentration  range is provided in Figure S4. In all systems, there is a delay in the onset of shrinkage in DNA chain length (Figure S4). This behavior is consistent with the observations and discussions for \textcolor{red}{\textit{ds 17}} in Figure \ref{fig2}. Importantly, the onset of these changes occurs at progressively higher concentrations of [$BMIM$] acetate as the proportion of \textit{ds} nucleotides increases. Specifically, the DNA shrinkage onset concentrations for systems with \textit{ds 17}, \textit{ds 32}, \textit{ds 53}, and \textit{ds 90} are approximately $60~mM$, $80~mM$, $100 ~mM$, and $100~mM$, respectively. This trend can be explained by considering that increase in dsDNA promotes increased structural rigidity and higher persistence length. Thus, higher concentration of [$BMIM$] acetate is required to displace $Na^+$ ions and achieve the critical threshold for inducing changes in DNA chain length. As seen in Figure \ref{fig3}b, the estimates of first power law exponent, $\alpha_1$, increases with increasing composition of double-stranded DNA. Specifically, estimates of \textit{0.14}, \textit{0.15} \textit{0.19}, and \textit{0.23} were obtained for \textit{ds 17}, \textit{ds 32}, \textit{ds 53}, and \textit{ds 90} respectively. However, in the high [$BMIM$] acetate concentrations regime, the slopes, $\alpha_2$, were observed to progressively decrease with increasing composition dsDNA with estimates of \textit{0.22}, \textit{0.18}, \textit{0.13}, and \textit{0.05} for \textit{ds 17}, \textit{ds 32}, \textit{ds 53}, and \textit{ds 90} respectively. A larger composition of double-stranded DNA (dsDNA) leads to an earlier transition to the stronger interaction  $\alpha_2$ regime, indicating that longer dsDNA chains are the cause of this effect. The onset concentrations were found to be $\sim$ $250~mM$, $250~mM$, $150~mM$, and $125~mM$ for \textit{ds 17}, \textit{ds 32}, \textit{ds 53}, and \textit{ds 90} respectively.\\
To understand the dependence of $\alpha_1$ and $\alpha_2$ on dsDNA composition, let us first consider the response of \textit{ds 17} at low salt, where $ds~ to~ ss$ ratio is 17 $\%$. In this region, we measure $\alpha_1$ $\approx~0.14$ similar to monovalent inorganic $NaCl$ salt, indicative of the dominance of electrostatic interactions\cite{srivastava2022effect}. As the IL concentration increases, groove binding mechanism and electrostatic interactions, both comes into play and we measure an $\alpha_2$ $\approx~0.22$ as shown and discussed earlier Figure \ref{fig2}. The \textit{ds 90} which has $\sim~90~\%$ dsDNA and for simplicity it can be assumed that interaction between the IL and DNA to occur exclusively with dsDNA. In this system the onset of groove binding can be seen at low salt as it is easier for IL to form groove binding with major and minor grooves of dsDNA. This leads to faster decrease in DNA chain length and hence higher estimate of $\alpha_1$ (Figure \ref{fig3}b) is measured. For \textit{ds 32} and \textit{ds 53}, the cumulative effect of electrostatic and groove binding interactions results in the estimates of $\alpha_1$ and $\alpha_2$ which lie between \textit{ds 17} and \textit{ds 90} (Figure \ref{fig3}b). In summary we measure a systematic increase in the contribution from groove binding interactions with increase in \textit{ds} composition of DNA, whereas at large \textit{ss} composition the nanoscale interactions primarily dominated by electrostatics at low [$BMIM$] acetate concentration.\\
\begin{figure*}[!ht]
\centering
\includegraphics[width={1\textwidth}]{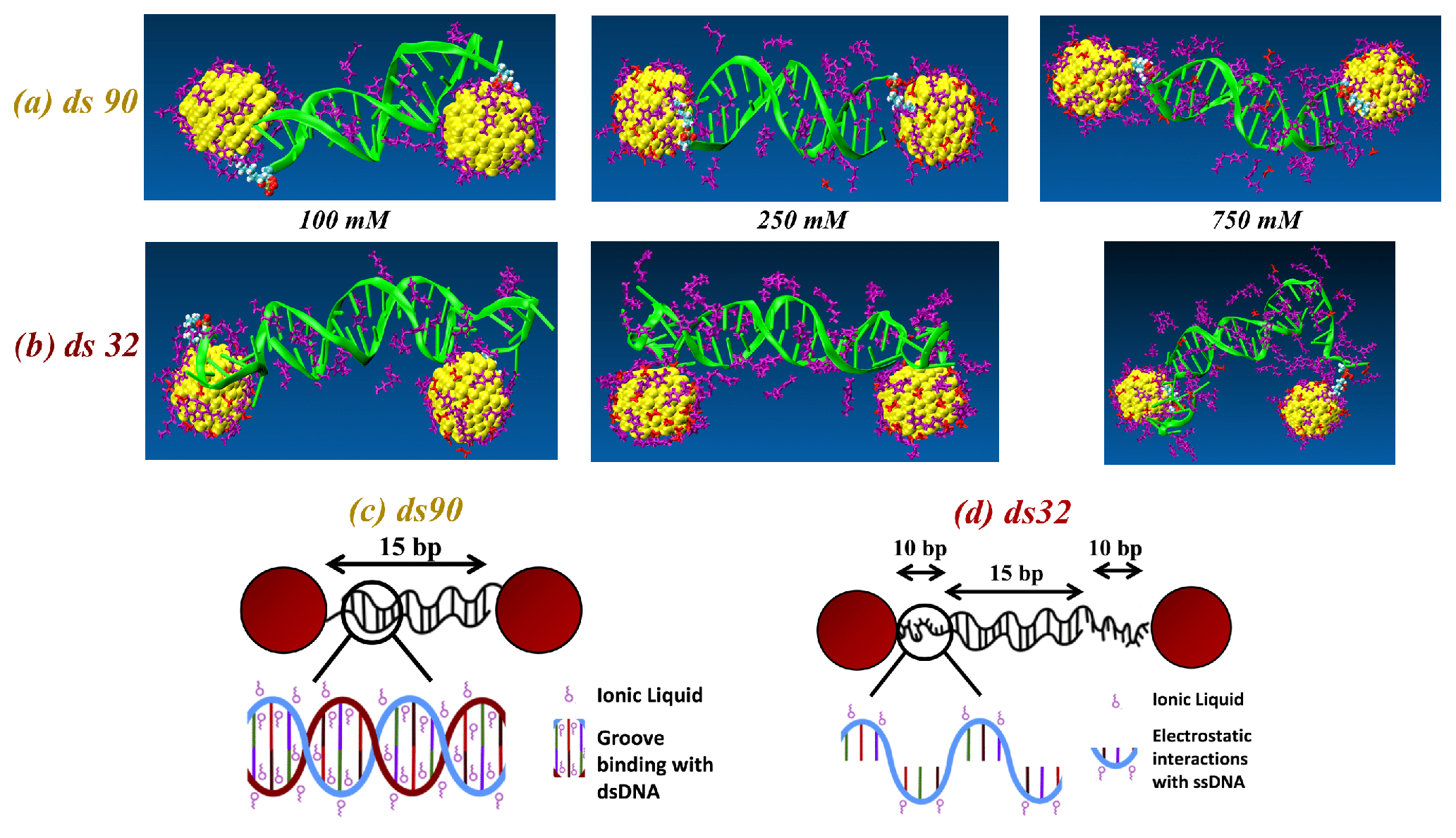}  
\caption{ \textit{Representative instantaneous snapshots of DNA-grafted AuNPs at three different ionic liquid (IL) concentrations from the final frames of 200 ns all-atom molecular dynamics (MD) simulations. Based on the percentage of paired dsDNA bases, the two systems are defined as follows: (a) dsDNA-rich (\textit{ds 90}) system, which exhibits pronounced groove binding of IL cations even at lower ionic strengths, and (b) ssDNA-rich (\textit{ds 32}) system, where DNA brush length undergoes significant compaction at higher ionic strengths due to dominant electrostatic interactions between IL cations and the flexible single-stranded segments. These distinct behaviors are further illustrated in the schematics shown in - (c) dsDNA-rich (\textit{ds 90}) and (d) ssDNA-rich (\textit{ds 32}) systems, respectively.}}
\label{fig4}
\end{figure*}
To validate the hypothesized interaction mechanisms observed in our experiments, we performed atomistic MD simulations on two systems: one predominantly consists of unpaired ssDNA nucleotides and the other entirely of dsDNA nucleotides with complementary Watson-Crick (WC) base pairs, corresponding to the experimentally studied systems \textit{ds 32} (32\% double stranded nucleotides) and \textit{ds 90} (90\% double stranded nucleotides), respectively. MD simulations were conducted at varying concentrations of ionic liquid (IL). Instantaneous snapshots of the configurations at the end of 200 ns long MD simulations are shown in Figures \ref{fig4}a-b.

The final morphology of \textit{ds 90} system indicates that the DNA end-to-end chain length $L_e$ (defined as the DNA brush length $H$ in experiment, with $L_e$=$2H$) remains nearly stable across all three IL concentrations (i.e, $100$, $250$, and $750 mM$), in absence of DNA coiling on the spherical AuNP surface and minimal DNA bending. This stability arises from IL binding to the dsDNA grooves, which stiffens the duplex as shown in Figure \ref{fig4}a. In contrast, \textit{ds 32} system exhibits progressive compaction of $L_e$ with increasing IL concentration due to the bending of flexible ssDNA-rich unpaired nucleotides, dominated by electrostatic interactions as shown in Figure \ref{fig4}b. In case of \textit{ds 90} system, schematics in Figure \ref{fig4}c highlight that ILs predominantly bind to DNA grooves at all three IL concentrations (see Figure \ref{S5} in the supporting information for detailed structural characterization of IL binding within the DNA grooves at $750 mM$). Whereas, in ssDNA-rich \textit{ds 32} system, electrostatic interactions between cationic ILs and unpaired DNA nucleotides dominate at higher IL concentrations, where groove binding remains comparatively weaker as described in the schematics in Figure \ref{fig4}d.
\begin{figure*}[!ht]
\centering
\includegraphics[width={1\textwidth}]{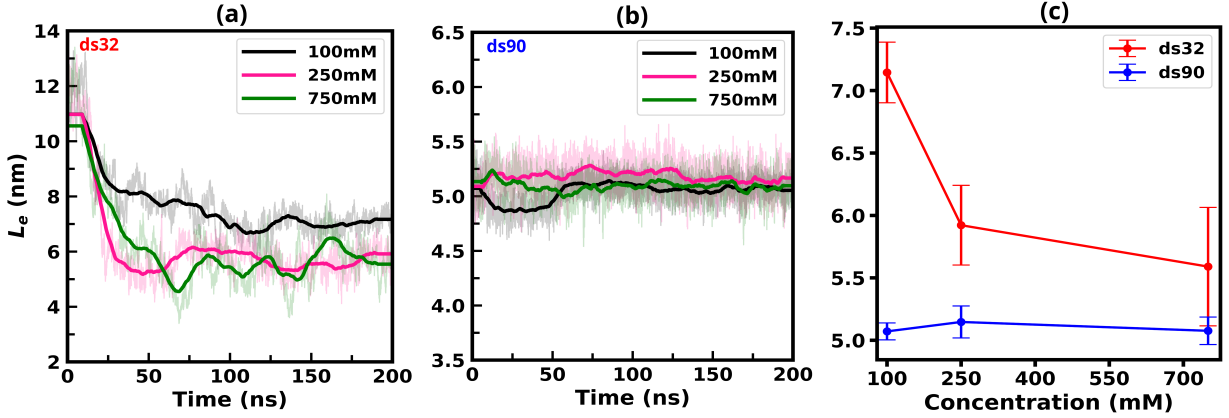}  
\caption{ \textit{ End-to-end distance ($L_e$) of the DNA chains, calculated from the 200 ns MD simulation trajectory, shows that with increasing ([$BMIM$] acetate) IL concentration, DNA chain length decreases, strongly modulated by the relative ssDNA and dsDNA compositions. (a) In ssDNA-rich systems (i.e.,\textit{ds 32}), faster chain compaction is observed over time, with increasing IL concentrations. (b) The dsDNA-rich \textit{ds 90} exhibit significantly less reduction in $L_e$ due to the increased chain stiffness arising from DNA groove bindings as compared to \textit{ds 32}. (c) Comparative variation of $L_e$ across three different ILs concentrations for both ssDNA- and dsDNA-rich systems, highlighting the distinct compaction regimes. }}
\label{fig5}
\end{figure*}

To elucidate the morphological characteristics and tuning of DNA brush lengths upon IL binding, we computed $L_e$ as a function of time and for different IL concentrations ($C_s$), as illustrated in Figure \ref{fig5}a-c. Our analysis of $L_e$ confirms that \textit{ds 90} undergoes only minor structural changes with increasing IL concentration as depicted in Figure \ref{fig5}b-c, consistent with the experimental data. In contrast, the \textit{ds 32} system exhibits a faster decrease in $L_e$ than the \textit{ds 90} system at higher IL concentrations, as shown in Figure \ref{fig5}a-c. This difference reflects the balance between groove binding, which stiffens dsDNA, and strong electrostatic interactions, which drive compaction of ssDNA-rich part. The calculated $\alpha$ (power law exponent in $H~\sim C_s^{-\alpha}$, as discussed in details in the experimental part) from MD simulations was found to be $\sim 0.15$ for \textit{ds 32} and $\sim 0$ for \textit{ds 90}, similar to what is observed experimentally. The simulation results are consistent with our experimental findings and provide critical molecular insights into the fundamental mechanism of DNA-IL interactions.

\begin{figure*}[!ht]
\centering
\includegraphics[width={1\textwidth}]{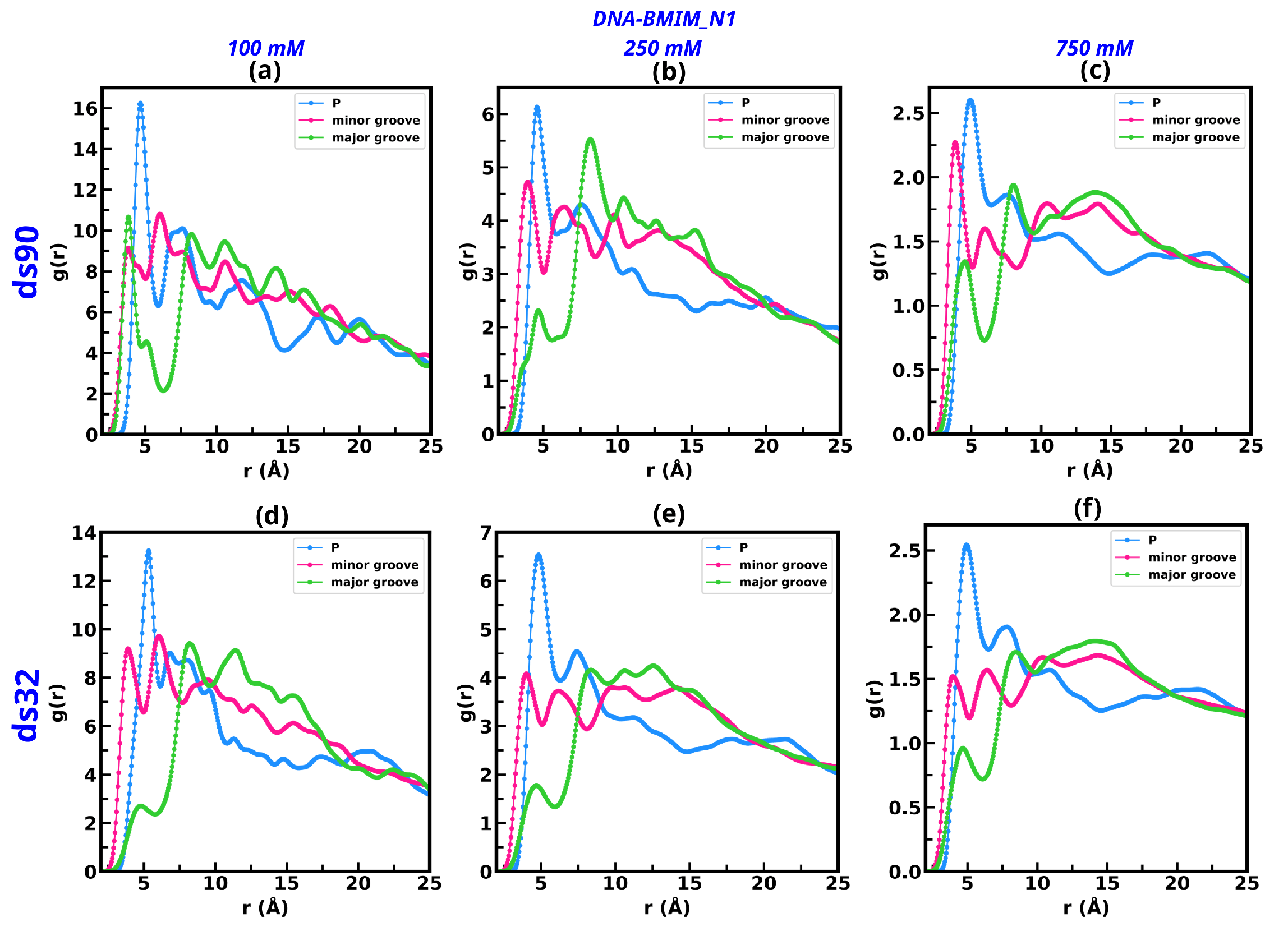}  
\caption{ \textit{The site-specific radial distribution functions (RDFs) of [$BMIM$] cationic N1 atoms around DNA phosphate group, minor groove, and major groove atoms are presented. The calculated RDFs are shown for dsDNA-rich \textit{ds 90} systems (a-c) and ssDNA-rich \textit{ds 32} systems (d-f) at IL concentrations of 100, 250, and 750 mM. RDFs were calculated using P atoms for phosphate groups. N3 and O2 atoms represent the minor groove electronegative sites. N7, N6, and O4 atoms were used for the major groove  electronegative sites. Colour coding is as follows: phosphate P-N1 (blue lines), minor groove-N1 (pink lines), and major groove-N1 (green lines).}}
\label{fig6}
\end{figure*}

In order to probe atomistic insights into DNA softening with increasing IL concentration, we analyzed site-specific interaction characteristics between ILs and DNA chains by calculating radial distribution functions (RDFs) for different  DNA regions. Figure \ref{fig6}a-c represents the distribution function of cationic head group N1 atoms of [$BMIM$] IL around the phosphate backbone, minor and major groove of DNA at \textit{100}, \textit{250}, and \textit{750 mM} IL concentrations. In dsDNA-rich \textit{ds 90} system (upper panel), the P-N1 RDF peak intensity remains the highest at all concentrations (Figure \ref{fig6}a-c, blue lines), reflecting dominant electrostatic interactions between the cationic imidazolium head group N1 atoms and the negatively charged P atoms of the phosphate backbone. However, the minor groove-N1 RDF peak (located at $\sim$4.2 {\AA}) intensity increases systematically with IL concentration, indicating enhanced accumulation and intrusion of IL cations into the minor groove at higher concentrations (\textit{750 mM}) as shown in Figure \ref{fig6}c (pink line). The increasing minor groove-N1 RDF peak heights relative to P-N1 indicate that minor groove association becomes progressively more significant at higher IL concentrations. The major groove exhibits the lowest RDF peak intensity at longer distances, $\sim$8 {\AA}, with increasing IL concentration (Figure \ref{fig6}a-c, green lines).

In ssDNA-rich \textit{ds 32} system (Figure \ref{fig6}d-f, lower panel), RDFs were calculated for the central 15 bp dsDNA, to make results consistent with the \textit{ds 90} system. In \textit{ds 32} system, the P-N1 interaction dominates at all concentrations with higher RDF peaks. In contrast to \textit{ds 90}, here we observed that the minor groove binding is evident at low IL concentration (\textit{100 mM}) but decreases at higher concentrations. Consequently, the minor groove-N1 RDF peak intensity reduces relative to the backbone P-N1 peak with increasing IL concentration from \textit{100} to \textit{750 mM}, where ILs instead interact strongly with phosphate groups via electrostatic interactions. The major groove-N1 RDF does not show any sharp peaks at higher IL concentrations, indicating that backbone P-N1 electrostatics govern IL binding in the ssDNA-rich \textit{ds 32} system.

The reduced RDF peak at higher IL concentrations arises from the enhanced clustering of the IL ions (i.e., [$BMIM$] cations and acetate anions), which prevents the formation of a well-defined solvation shell around the DNA molecule. At low IL concentration (100 mM), acetate anions preferentially form small water-mediated clusters, leaving [$BMIM$] cations more available for closer interactions with DNA regions, resulting in higher RDF peaks \cite{manna2020structure}. In contrast, at higher concentrations (250 and 750 mM), the stronger attraction between [$BMIM$] cations and acetate anions is retained, reducing the effective availability of the [$BMIM$] cations near DNA and leading to reduced RDF peak intensities.

In summary, RDF results indicate that IL shows strong minor groove binding to dsDNA-rich \textit{ds 90}, thereby enhancing DNA rigidity across all concentrations. In contrast, for \textit{ds 32}, groove binding weakens with increasing IL concentration, while electrostatic interactions become dominant. These findings are consistent with our experimental observations. The concentration-dependent RDFs between DNA and the cationic N1 atom, as well as between DNA and the imidazolium ring center-of-mass (COM), show similar trends of groove binding and electrostatic interactions, as discussed in more detail in Figures \ref{S6}-\ref{S9} of the Supporting Information.

\begin{figure*}[!htbp]
\centering
\includegraphics[width={1\textwidth}]{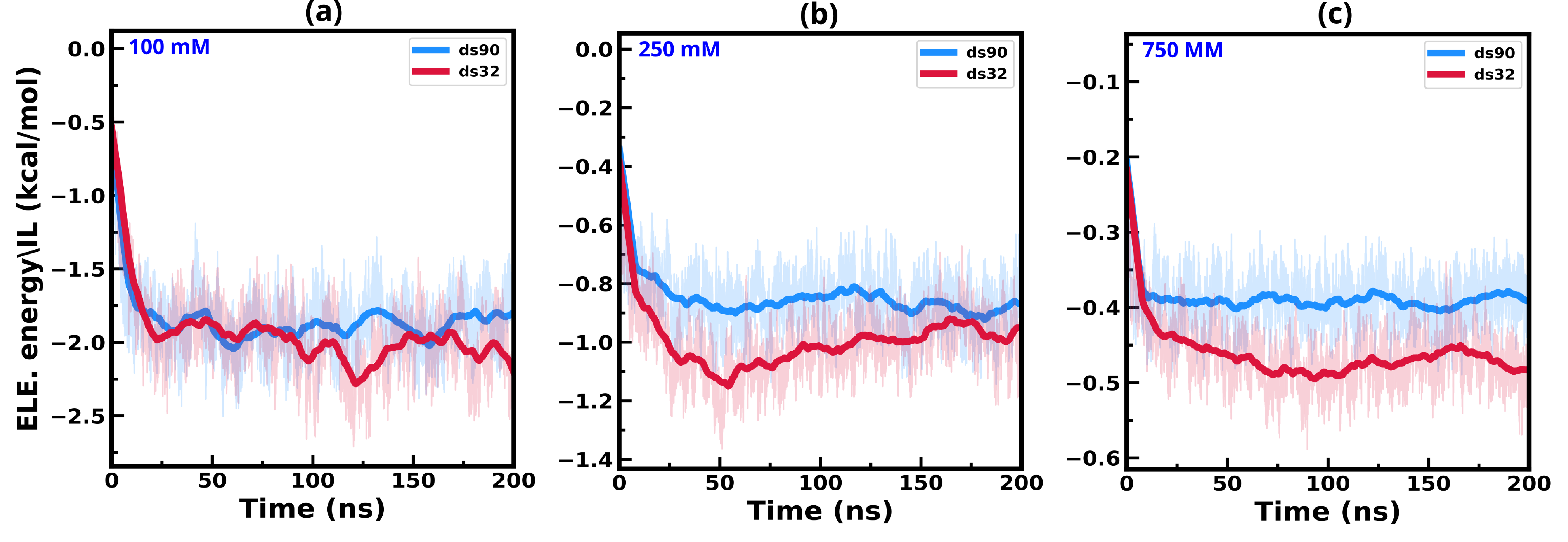}  
\caption{ \textit{Non-bonded electrostatic interaction energy-per IL between ILs and DNA chains grafted onto AuNPs, calculated for the dsDNA-rich (\textit{ds 90}) and ssDNA-rich (\textit{ds 32}) systems at IL concentrations of (a) 100 mM, (b) 250 mM, and (c) 750 mM. The color coding is as follows:: blue corresponds to the\textit{ds 90} system, and red corresponds to the \textit{ds 32} systems.}}
\label{fig7}
\end{figure*}

From the RDF analyses, we observed a dominant contribution of electrostatic interactions for IL cations binding to unpaired ssDNA bases and backbones in \textit{ ds 32}. Hence, we calculated non-bonded electrostatic interaction energies per IL, between cationic ILs and DNA nucleotides across the simulation time frames for both \textit{ds 32} and \textit{ds 90} systems at \textit{100 mM}, \textit{250 mM}, and \textit{750 mM} concentrations, as shown in Figures \ref{fig7}a-c. Electrostatic interaction energy values confirm the trends of much stronger IL-DNA interactions in \textit{ ds 32} than \textit{ ds 90}, explaining its sharp compaction of $L_e$, primarily driven by electrostatic interactions and enrichment of the unpaired ssDNA-bases (red curve in Figure \ref{fig7}a-c).

To characterize of electrostatic environment of DNA in ILs, we computed time-averaged 2D electrostatic potential maps obtained by solving Poisson’s equation using the final 20 ns of MD trajectories at \textit{100 mM}, \textit{250 mM}, and \textit{750 mM} IL concentrations, as shown in Figures \ref{S10}-\ref{S12}, see the supplementary information for details of the calculations. These potential maps provide insight into how the cationic ILs redistribute around the negatively charged dsDNA- and ssDNA-rich systems. In the dsDNA-rich \textit{ds 90} system, the helical structure and canonical charge distributions remain intact across all IL concentrations. The 2D electrostatic potential map shows negative potential regions along the phosphate backbone, and grooves are preserved, indicating that IL cations predominantly engage in groove binding without disrupting duplex integrity (Figure \ref{S12}, upper panel, SI).
In contrast, the ssDNA-rich \textit{ds 32} system shows pronounced compaction at $750 mM$ IL concentration. This is reflected in the 2D potential maps (Figure \ref{S12}, lower panel in SI), where the canonical helical charge distribution is no longer preserved. Instead, we observe broadened and expanded potential regions with partially neutralized negative charges, reflecting strong electrostatic binding of IL cations to unpaired bases. These results highlight the resilience of dsDNA (\textit{ds 90}) to IL perturbation, while ssDNA is highly susceptible to IL-driven compaction and charge neutralization (Figure \ref{S12}).
From these results, we conclude that the change in brush length of DNA chains is influenced by the relative composition of ssDNA-to-dsDNA bases in the presence of ILs. A higher proportion of paired WC bases in dsDNA favors groove binding, whereas the greater composition of unpaired bases in the ssDNA-rich system exhibits a strong concentration-dependent electrostatic interaction with ILs.

\section{Conclusion}
This study explores the morphological changes of DNA chains in imidazolium-based ionic liquid environments, revealing nanoscale interactions between DNA and IL. Small Angle X-ray Scattering experiments demonstrate that interactions for DNA chain nucleotides are influenced by IL concentration, leading to a reduction in the DNA chain length. The experiments further reveal that the interplay between electrostatic interactions and groove binding depends on the composition and concentration of IL and the composition of the DNA chain. For ssDNA, low IL concentrations primarily induce electrostatic interactions, as ssDNA is changed to dsDNA, groove binding becomes the dominant interaction, causing a faster reduction in chain length, even at low IL concentration. Molecular dynamics simulations support these findings, showing that IL interacts with the DNA backbone as well as with the minor grooves. The dominant interaction depends on DNA chain type (ssDNA or dsDNA-rich). For dsDNA-rich composition, the simulation confirms groove binding interactions into the minor grooves of DNA. This leads to increased stiffness at higher IL concentrations and slows the reduction in DNA chain length, which is consistent with both experimental and simulation analysis.\\

\section{Materials and Methods}
\subsection{Experimental Details}
Thiol-modified DNA chains and DNA linkers are purchased from Sigma Aldrich. The specific DNA sequences are provided in the supporting information (SI). Gold nanoparticles, AuNP, of diameter $\sim 15~nm$ are obtained from Ted Pella. ILs with purity $\sim$ 95 $\%$ and two different chain lengths: [$EMIM$] (1-ethyl-3-methylimidazolium) acetate ($C_8H_{14}N_2O_2$, $170.21~g/mol$) and [$BMIM$] (1-ethyl-3-butylimidazolium) acetate ($C_{10}H_{18}N_2O_2$, $198.26~g/mol$), are purchased from Sigma Aldrich. Structurally [$BMIM$] has four carbon chains and [$EMIM$] has two carbon chains attached to the third nitrogen atom. The molecular structures of [$BMIM$] and [$EMIM$] acetate, are shown in Figure \ref{S1}. IL solution of varying concentration in the range of $0.01~M$ to $1.5~M$ are prepared in $10~mM$ phosphate buffer.\\
In a typical experiment, gold nanoparticles are functionalized with 50-base single-stranded DNA (ssDNA) containing 15-base sticky ends, which enables them to connect with complementary particles through linkers, facilitating the formation of self-assembled structures. For AuNP functionalization, DNA and AuNPs are mixed in a molar ratio of 1000:1 and incubated at room temperature for 8 hours at constant pH of 7.4. A sodium chloride ($NaCl$) solution is gradually added over 8 hours until a final concentration ($C_s$) of 0.2 M is achieved. The mixture is then centrifuged to remove any excess DNA. Linkers of desired sequences are then added to facilitate assembly through Watson-Crick interaction. The samples are annealed for 2 hours few degrees above the DNA melting temperature to maximise the linker attachment. Further purification and removal of excess linkers are done through centrifugation. Equimolar complimentary DNA-grafted AuNPs are mixed, incubated for the formation of a self-assembled structure. Four systems, \textit{ds n}, where \textit{n} represents the percentage of dsDNA composition in DNA chain have been prepared by varying the ratios of double-stranded DNA (dsDNA) to single-stranded DNA (ssDNA). These ratios were 15:70 - \textit{ds 17}, 27:58 - \textit{ds 32}, 45:40 - \textit{ds 53}, and 77:8 - \textit{ds 90}. \par

Samples with DNA-AuNP aggregates at different ionic liquid concentrations are sealed in quartz capillaries and in solution structural changes are examined using the Small Angle X-ray Scattering. The SAXS measurements were conducted at the Complex Materials Scattering (CMS, 11-BM) beamline of the National Synchrotron Light Source II (NSLS-II) at Brookhaven National Laboratory. The scattered data were collected using a beam energy of 13.5 keV and beam size of 200 $\times$ 200 um with a Pilatus 2M area detector (Dectris, Switzerland). The detector, consisting of 0.172 mm square pixels in a 1475 $\times$ 1679 array, was placed five meters downstream from the sample position. The collected 2D scattering patterns were reduced to 1D scattering intensity, $I(q)$, by circular average. The $q$ is wave vector transfer, $q = (4\pi/\lambda) sin(\theta)$, where $\lambda$ = 0.9184 Å and $2\theta$ is the wavelength of the incident x-ray beam and the scattering angle, respectively. The schematic representation of the experimental setup is shown in Figure \ref{fig1}. The structure factor $S(q) = \frac{I(q)}{F(q)}$,  measured from the X-ray scattering intensity profiles. \textit{F(q)} is the nanoparticle form factor, obtained from measuring the scattering intensity of dispersed DNA–AuNPs. The form factor, \textit{F(q)} for DNA-AuNP of diameter $15~nm$ dispersed in aqueous medium is shown in Figure \ref{S2}.\par

\subsection{MD Simulation Details}
For the atomistic simulations, we have considered DNA chains grafted onto $2~nm$ (diameter) nm gold nanoparticles (AuNP) in two systems: one enriched with unpaired ssDNA bases and the other system predominantly composed of all paired dsDNA WC bases. The ssDNA-rich system was designed to mimic the $ds~32$ system, and the dsDNA-rich system was designed to mimic the $ds~90$, as per experimental designs. In the ssDNA-rich system, each AuNP was functionalized with a 25-base ssDNA chain via a thiol-linker\cite{kumar2012structure}. The outer 15 bases of each ssDNA chain form Watson-Crick (WC) base pairs with the complementary strand on the other AuNP, creating a self-assembled structure with 15 WC-paired bases in the center and 20 (10+10) unpaired ssDNA bases, grafted between the two AuNPs, as illustrated by the schematic in Figure \ref{fig4}d. In the dsDNA-rich system, each AuNP was grafted with a 15-base DNA strand that fully pairs with the complementary strand on the other AuNP, forming a complementary WC duplex without unpaired bases, as shown in Figure \ref{fig4}c. The ionic liquid 1-butyl-3-methylimidazolium acetate ([$BMIM$] acetate) was used to modulate the morphology of the DNA-grafted AuNPs.
In this study, MD simulations were carried out at three IL concentrations of 100, 250, and 750 $mM$. The initial DNA structures were built using the NAB\cite{macke1998modeling} code of the AMBER module. A thiol group served as a linker between the DNA strands and AuNPs, and the resulting DNA-thiol-AuNP conjugates were solvated in a water box using the TIP3P water model. To model the interaction of DNA atoms, we employed interaction parameters from the CHARMM27 force field \cite{mackerell1998all}. Water was modeled using TIP3P, as it is a well-established water model for DNA simulation. For the neutralizing ions, we have used Joung-Chetham ion parameters. The interactions involving Au atoms were described using Lennard-Jones (LJ) parameters, with $\sigma = 0.2620$ nm and $\epsilon = 22.13$ kJ.mol$^{-1}$, as well as the parameters for the alkylthiol linker were adopted from the literature \cite{hautman1989simulation}. For the ionic liquid (IL), 1-butyl-3-methylimidazolium acetate ([$BMIM$] acetate), we used interaction potential parameters obtained from previous studies \cite{sambasivarao2009development,doherty2017revisiting,canongia2006nanostructural}, which were developed based on the OPLS-AA force field framework.
All MD simulations were carried out in GROMACS \cite{hess2008gromacs}. Initially, the energy minimization was performed using the steepest descent algorithm. The systems were then allowed to equilibrate for 2 ns in the canonical NVT ensemble. To maintain the temperature at 300 K, a Berendsen thermostat with a coupling constant of 0.1 ps was applied for 5 ns. The positions of the Au atoms were restrained using a harmonic potential with a force constant of 1000 kJ.mol$^{-1}$.nm$^{-2}$. Subsequently, an additional 2 ns equilibration was performed in the isothermal–isobaric NPT ensemble, where the pressure was maintained at 1 atm using a Berendsen barostat with a coupling constant of 0.2 ps. To mimic bulk solvent conditions, periodic boundary conditions were applied in all three spatial directions. The LINCS algorithm was used to constrain all bonds involving hydrogens, allowing the use of a 2 fs integration time step. Long-range electrostatic interactions were treated using the Particle Mesh Ewald (PME) method, with a real-space cutoff of 1.2 nm. The same cutoff distance was applied for short-range van der Waals interactions. Following the equilibration, production run simulations were carried out in the NVT ensemble at 300 K for 200 ns. Similar simulation protocols have been employed in some of our previous studies on DNA-based systems \cite{mandal2024dynamics,kumbhakar2023utilization,singh2024dna,mandal2025mechanistic, mandal2025silico}.
\section{Acknowledgments}
SS acknowledges support from the CRS-UGC-DAE (CRS/2022-23/03/876), India. This research used resources of the Center for Functional Nanomaterials and the CMS beamline (11-BM) of the National Synchrotron Light Source II, both supported by U.S. DOE Office of Science Facilities at Brookhaven National Laboratory under Contract No. DE-SC0012704. AC acknowledges financial support from CRNTS IITB and IRCC IIT Bombay. This work was partially supported by DST/JSPS/P-389/2024(G). The authors thank SAIF-IITB for the SAXS facility. SM acknowledges SRF fellowship from CSIR, India. PKM thanks the Department of Science and Technology (DST), India, for financial support and the Science and Engineering Research Board (SERB) for financial and computational support through CRG/2021/003659.
\bibliography{REF}%

\clearpage
\clearpage
\section{Supporting Information}
\begin{table}[]
    \centering
    \begin{tabular}{|c|c|}
    \hline
\textbf{DNA system} & \textbf{DNA sequence} \\ 
 \hline
 Seq 1 & {\scriptsize $HSC_6H_{12}$-TTTTTTTTTTTTCGTTGGCTGGATAGCTGTGTTCTTAACCTAACCTTCAT}\\
 \hline
 Seq 2 & {\scriptsize $HSC_6H_{12}$-TTTTTTTTTTTTCGTTGGCTGGATAGCTGTGTTCTATGAAGGTTAGGTTA}\\
 \hline
 linker (ds 32) & {\scriptsize AGAACA} \\
 \hline
 linker (ds 53) & {\scriptsize AGAACACAGCTATCC} \\
 \hline
 linker (ds 90) & {\scriptsize AACACAGCTATCCAGCCAACGAAAAAAAAAA}\\
 \hline
    \end{tabular}
    \caption{The DNA sequence design (5’ to 3’) for systems presented in the paper. $HSC_6H_{12}$ represents the thiol modification. Seq 1, Seq 2 , ds32, ds53, and ds90 have 50, 50, 6, 15, and 31 bases respectively. Outer 15 bases for Seq 1 and Seq 2 are designed to form the Watson Crick Base pairing to form a self-assemble structure.}
    \label{Table S1}
\end{table}

\setcounter{figure}{0}
\renewcommand{\figurename}{Figure }
\renewcommand{\thefigure}{S\arabic{figure}}
\begin{figure}
    \centering
\includegraphics[keepaspectratio]{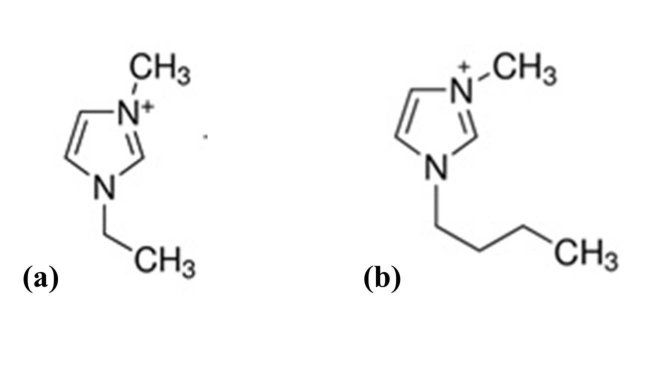}
    \caption{ \textit{Molecular structure of a)$[EMIM]^+$ b)$[BMIM]^+$. $[EMIM]^+$ has two carbon chains attached to the 3’ nitrogen whereas $[BMIM]^+$ has 4 carbon chain attached to the 3’ nitrogen.}}
    \label{S1}
\end{figure}

\begin{figure}
    \centering
\includegraphics[width=\textwidth,height=\textheight,keepaspectratio]{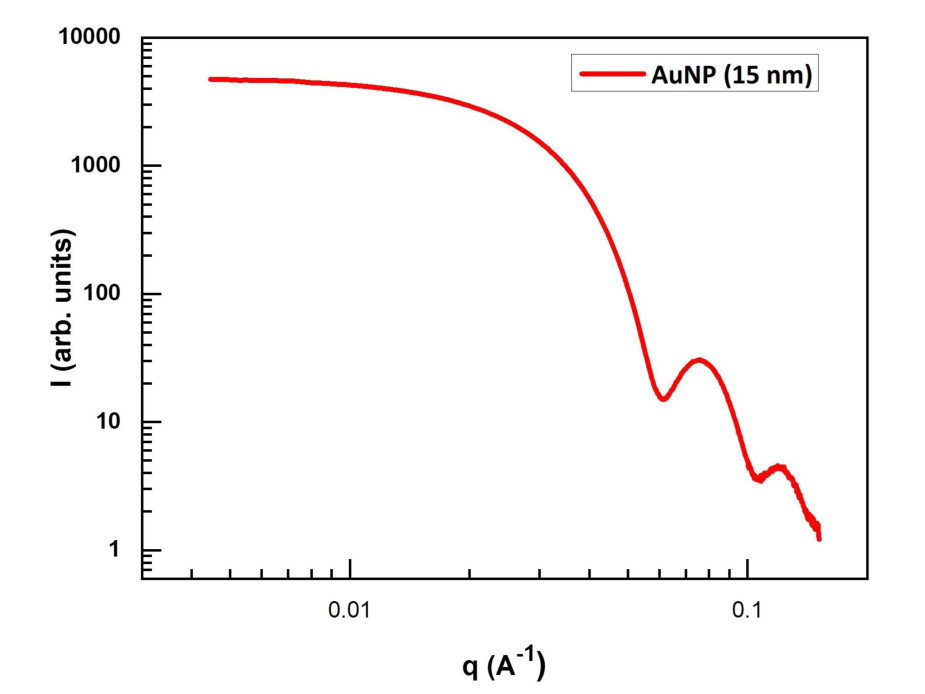}
    \caption{ \textit{Small Angle X-Ray scattering for 15 nm gold nanoparticle.}}
    \label{S2}
\end{figure}

\begin{figure}
    \centering
\includegraphics[width=\textwidth,height=\textheight,keepaspectratio]{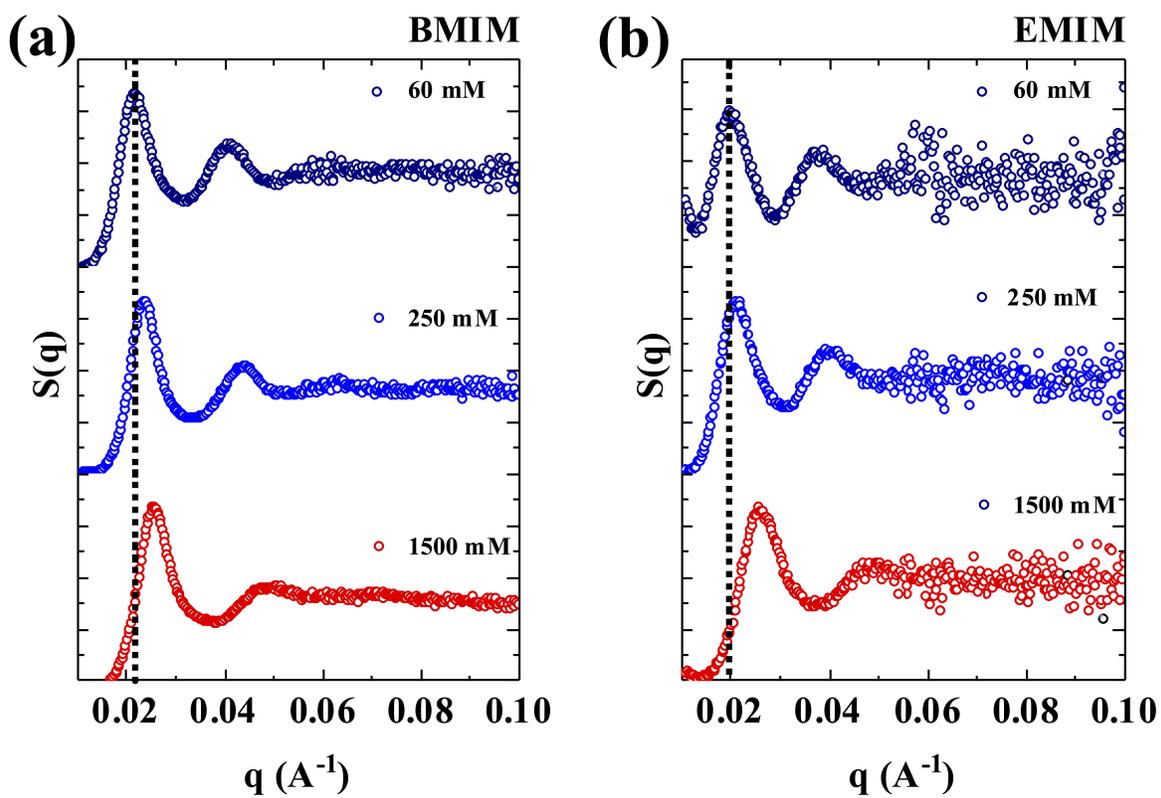}
    \caption{ \textit{Measured structure factor for ds 17 for a) [BMIM] acetate b) [EMIM] acetate. The structure factor peak shifts towards higher q, indicating the shrinkage of DNA chain length.}}
    \label{S3}
\end{figure}





\begin{figure}
    \centering
\includegraphics[width=\textwidth,height=\textheight,keepaspectratio]{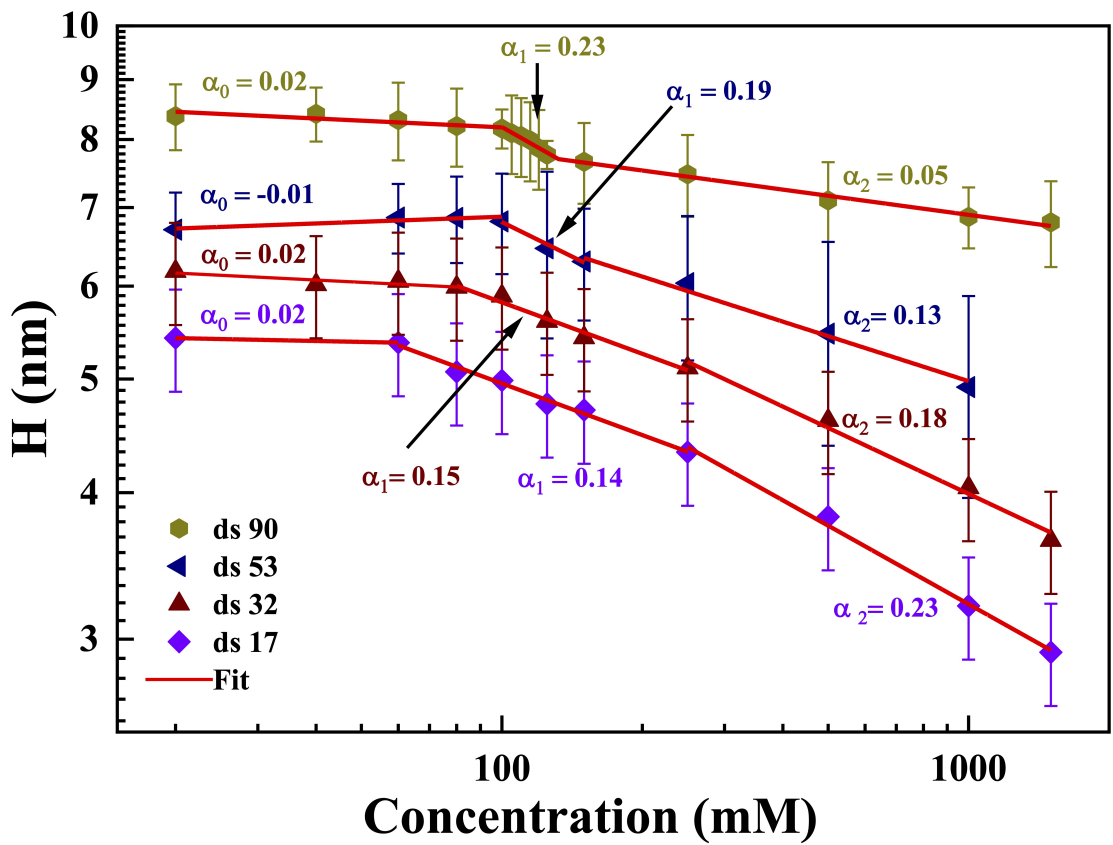}
    \caption{ \textit{Variation of H with increasing the [Bmim] acetate by varying the dsDNA to ssDNA ratio. ds17, ds32, ds53, and ds90 have a dsDNA to ssDNA ratio of 15:70, 27:58, 45:40, and 77:8, respectively.}}
    \label{S4}
\end{figure}

\section{DNA Groove Binding for 750 mM IL Concentration}
\begin{figure}[H]
    \centering
\includegraphics[width={1\textwidth}]{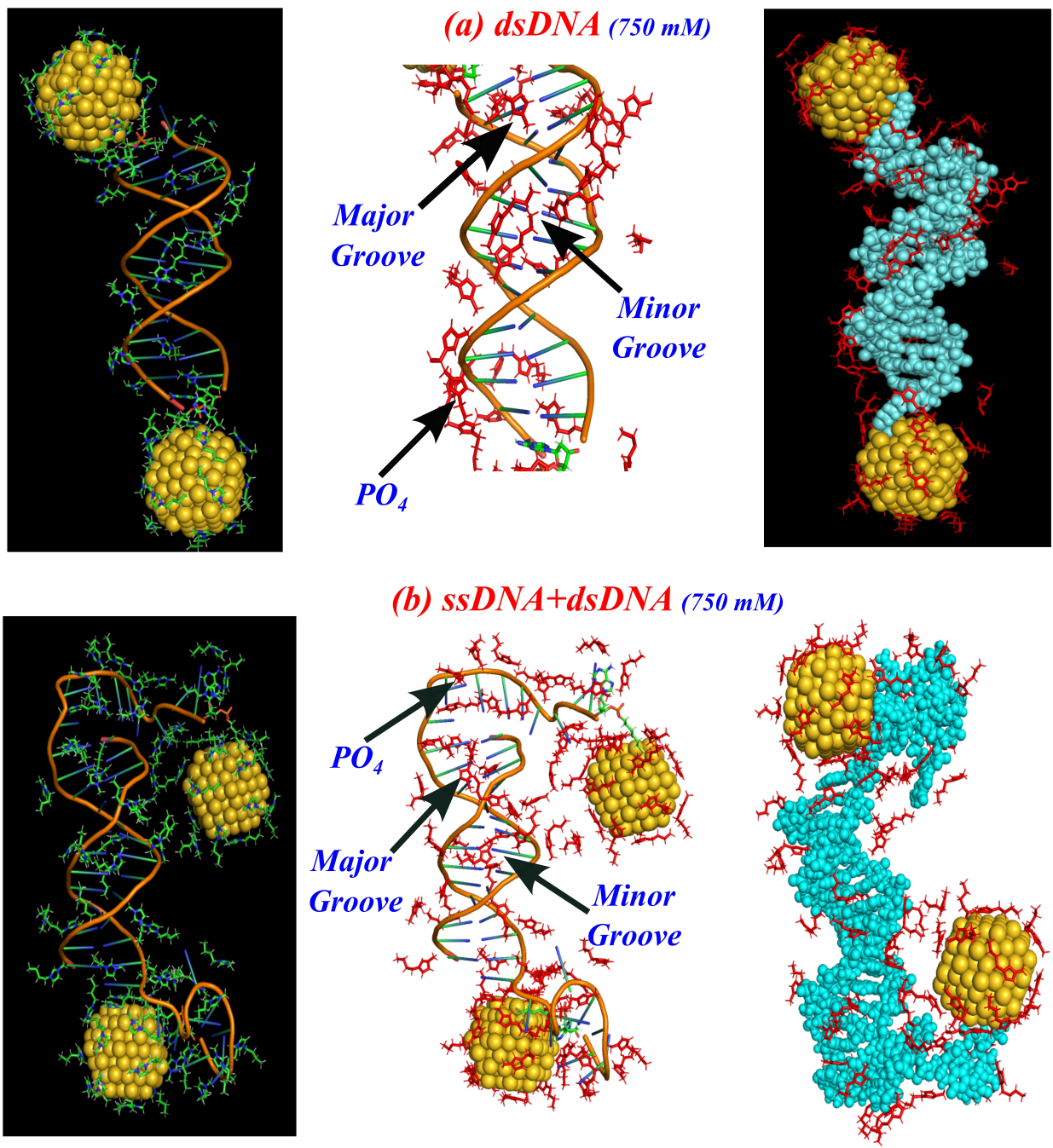}
    \caption{ \textit{Representative illustrations of DNA major/minor groove binding by [BMIM]acetate ILs at 750 mM, the concentration, where IL interactions are most prominent. (a) In dsDNA-rich systems (\textit{ds 90}), cationic ILs preferentially bind to the phosphate backbone and minor grooves, with comparatively weaker binding at the major groove. (b) However, in ssDNA-rich systems (\textit{ds 32}), IL binding is largely confined to the backbones of ssDNA parts, due to dominant electrostatic interactions rather than groove bindings.}}
    \label{S5}
\end{figure}

\section{Concentration-dependent Radial Distribution Functions}
\begin{figure}[H]
    \centering
\includegraphics[width={1\textwidth}]{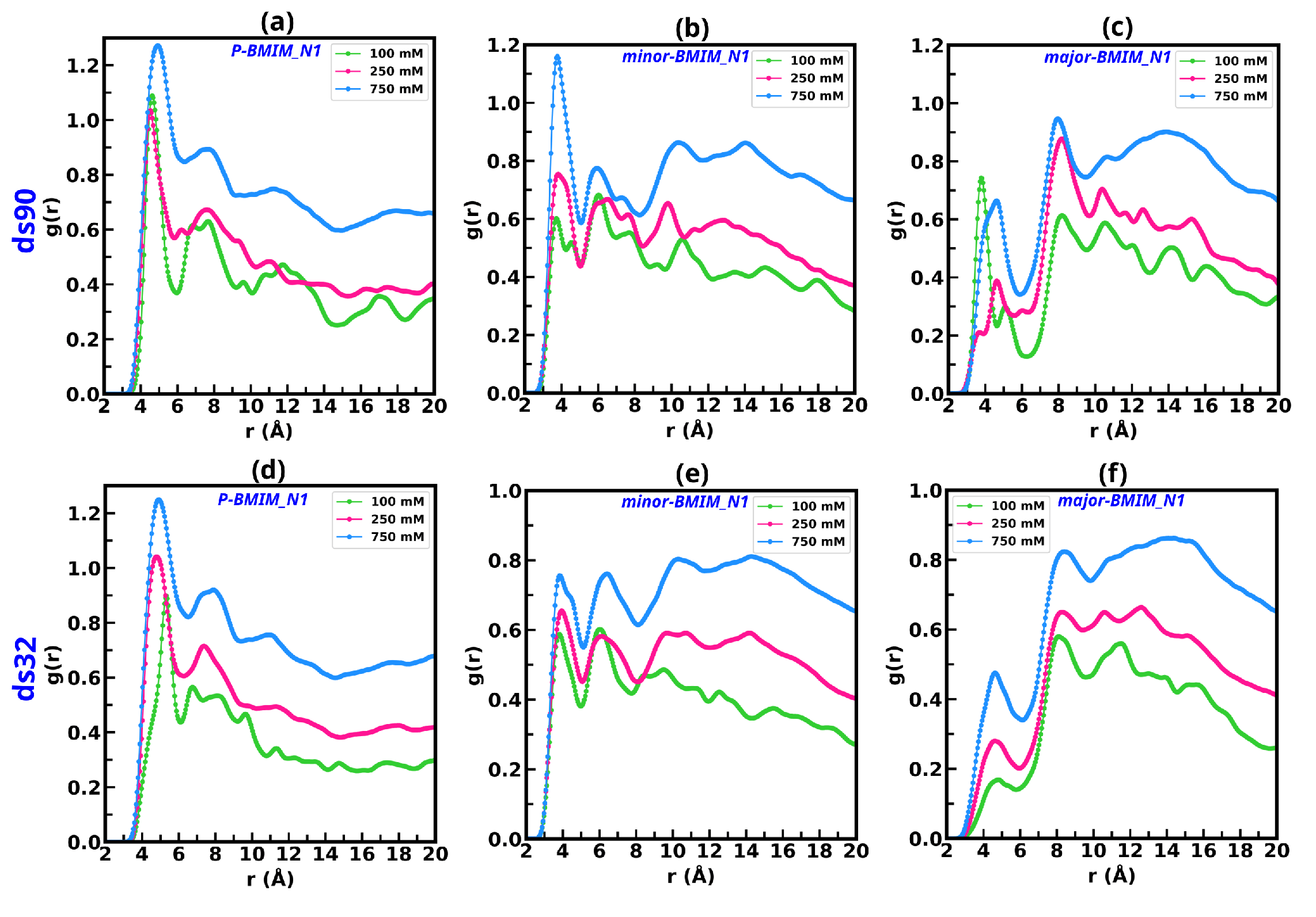}
    \caption{ \textit{Concentration-dependent site-specific radial distribution functions (RDFs) of the [$BMIM$] imidazolium cationic N1 atoms around different DNA regions. For dsDNA-rich \textit{ds 90} system (upper panel): (a) phosphate backbone (left panel), (b) minor groove (middle panel), and (c) major groove (right panel). For the ssDNA-rich \textit{ds 32} system (lower panel), RDFs correspond to the (d) phosphate backbone, (e) minor groove, and (f) major groove. IL concentrations are 100 mM (green), 250 mM (pink), and 750 mM (blue). All RDFs are normalized by the IL number density.}}
    \label{S6}
\end{figure}

We have computed concentration-dependent site-specific RDFs by multiplying the normalized RDF by the corresponding bulk IL number density for various IL concentrations (e.g. 100, 250, and 750 mM). These RDF now directly reflect the local population of IL cations around DNA regions. We observe monotonic increase in the cationic N1 peak intensity with IL concentrations for P backbone (left panel), minor groove (middle panel), and major groove (right panel) atoms in both \textit{ds 90} (Figure \ref{S6}.a-c) and \textit{ds 32} (Figure \ref{S6}.d-f) systems.  

For the \textit{ds 90} system (upper panel), the minor groove-N1 RDF peak exhibits an approximately 2-fold increase in intensity from 100 to 750 mM (Figure \ref{S6}.b), whereas the backbone P-N1 peak increases by $\sim$1.2-fold over the same concentration range (Figure \ref{S6}.a). The preferred localization distances remain essentially unchanged, with IL cations localized at $\sim$4.2 {\AA} from minor groove atoms (N3, O2) and $\sim$5 {\AA} from backbone P atoms. The sharp increase of the minor groove RDF peak intensity indicates enhanced groove-specific accumulation/binding at higher concentrations (750 mM) in the dsDNA-rich \textit{ds 90} system. 

In contrast, for the ssDNA-rich \textit{ds 32} system (lower panel), the sharp enhancement of the minor groove-N1 peak is absent (Figure \ref{S6}.e). Instead, the P-N1 peak shows a $\sim$1.5-fold increase with concentration from 100 to 750 mM (Figure \ref{S6}.d), indicating that electrostatic interactions with the phosphate backbone dominate IL binding in \textit{ds 32} system. At all IL concentrations, the major groove-N1 peaks (Figure \ref{S6}.c and \ref{S6}.f) are less pronounced than those of the minor groove. Although their intensities increase with concentration in both systems, minor groove binding is clearly more significant in \textit{ds 90}, whereas backbone P-N1 electrostatic interactions govern IL association in \textit{ds 32}.

In order to investigate the interaction between DNA base pairs and the imidazolium ring COM of the IL, we have computed site-specific RDFs between: (a) COM of the ring of DNA bases and the COM of the imidazolium ring, (b) DNA backbone P, major and minor groove regions, and COM of the imidazolium ring and (c) DNA backbone P, major and minor groove regions, and COM of imidazolium ring (concentration-dependent).
\begin{figure}[H]
    \centering
\includegraphics[width={1\textwidth}]{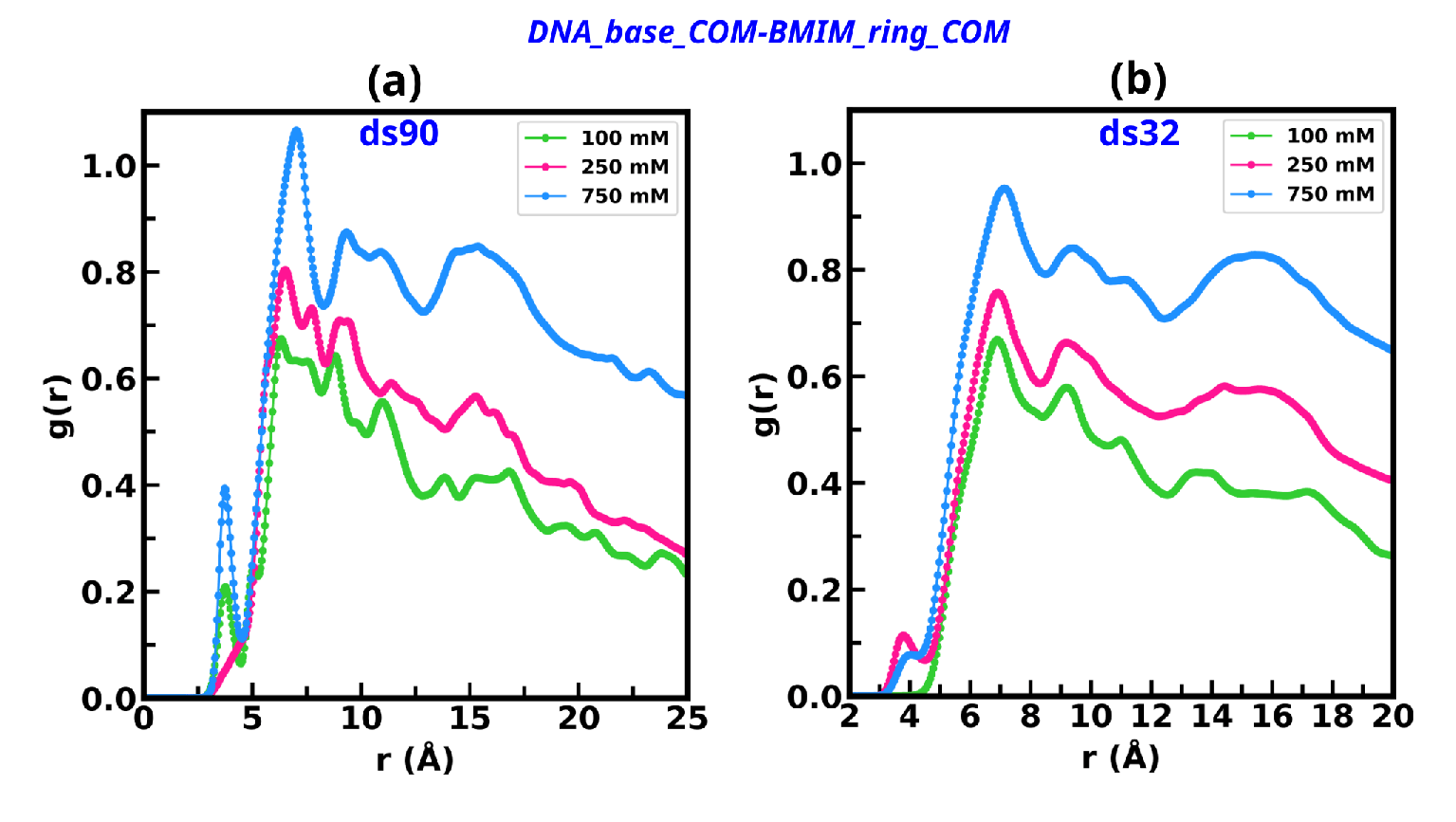}
    \caption{ \textit{Concentration-dependent radial distribution function (RDF) between COM of the ring of DNA bases and COM of the [BMIM] imidazolium ring for (a) \textit{ds 90} and (b) \textit{ds 32} system. The \textit{ds 90} has all paired DNA bases, and ssDNA-rich \textit{ds 32} has some unpaired bases. For all RDF calculations, we have used the middle 15 paired bases for the \textit{ds 32} system to ensure results are consistent with those of the \textit{ds 90} system. The IL concentrations are 100 mM (green line), 250 mM (pink lines), and 750 mM (blue lines).  All the RDFs are scaled by the IL number density.}}
    \label{S7}
\end{figure}

Figure \ref{S7} shows that, for \textit{ds 90}, the first RDF peak between COM of the ring of DNA bases and COM of the imidazolium ring, appears at $\sim$3.4 {\AA}, followed by a second peak near $\sim$7 {\AA}, with a pronounced increase in peak intensity as IL concentration increases from 100 to 750 mM (Figure \ref{S7}.a). This behavior indicates enhanced $\pi$-$\pi$ stacking and short-range association of the imidazolium ring with DNA bases. In contrast, \textit{ds 32} does not exhibit a prominent first peak at $\sim$3.4 {\AA}, and the second-shell ($\sim$7 {\AA}) RDF peak intensity is weaker than in \textit{ds 90} at higher IL concentrations, suggesting reduced base-stacking or groove binding interactions in the ssDNA-rich \textit{ds 32} system (Figure \ref{S7}.a). 

\begin{figure}[H]
    \centering
\includegraphics[width={1\textwidth}]{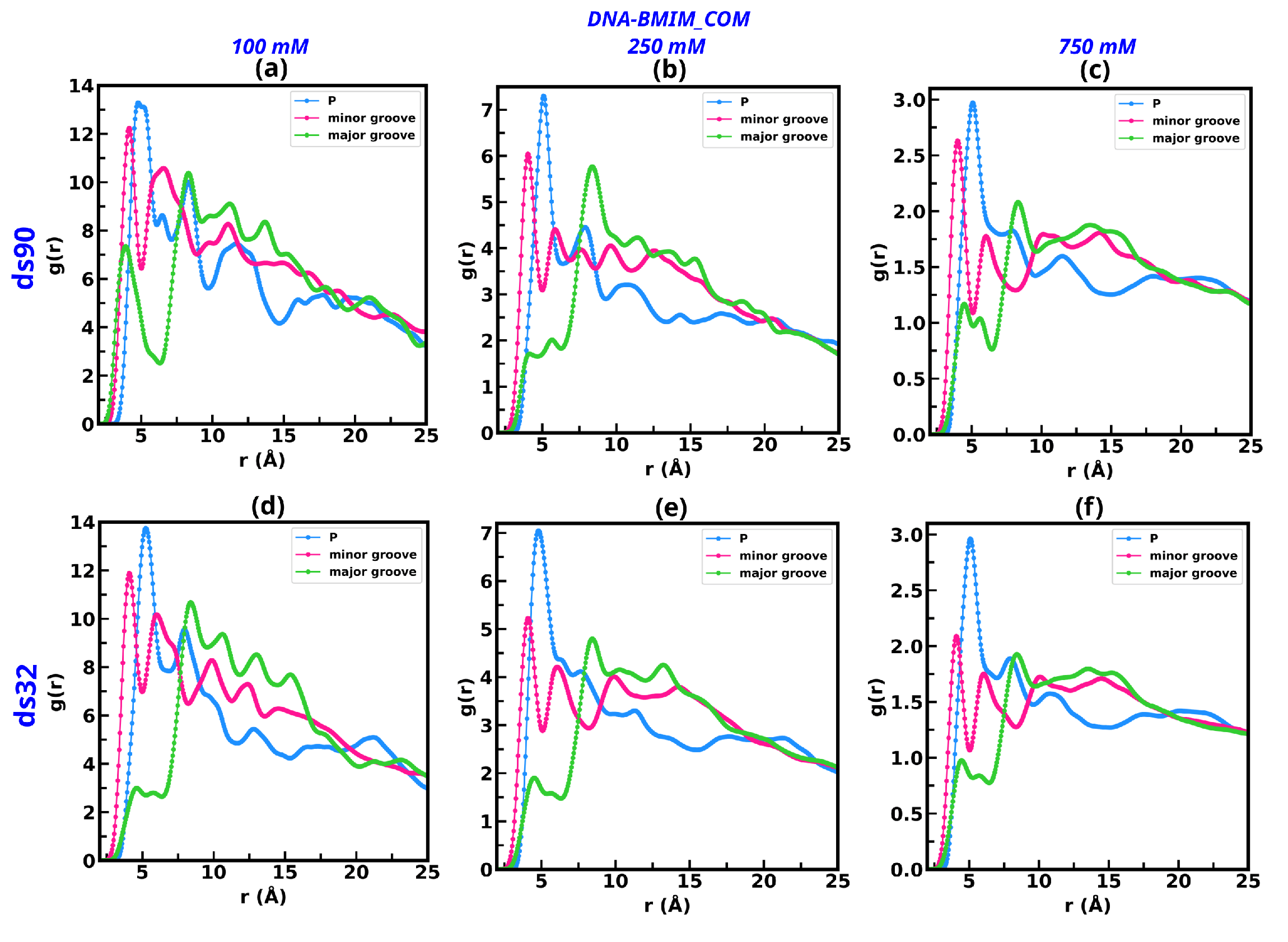}
    \caption{ \textit{The site-specific radial distribution functions (RDFs) for COM of the [BMIM] imidazolium ring around backbone P, minor, and major groove atoms of DNA duplexes in different IL concentration solutions. RDF calculations are performed using the N7, N6, and O4 atoms for the electronegative sites in the major groove. For the minor groove electronegative sites, we have used N3 and O2 atoms. P atoms were used for the phosphate groups. The upper panel represents RDF for the dsDNA-rich \textit{ds 90} system at different IL concentrations of (a) 100 mM, (b) 250 mM, and (c) 750 mM. The lower panel IL concentrations are (d) 100 mM, (e) 250 mM, and (f) 750 mM for the ssDNA-rich \textit{ds 32} system.}}
    \label{S8}
\end{figure}

RDFs in Figure \ref{S8}, provide evidence that IL imidazolium ring COM occupy space near the DNA backbone and minor groove regions. In \textit{ds 90} (upper panel), at 100 mM of IL concentration, the minor groove exhibits a strong first-shell RDF peak intensity at $\sim$4 {\AA}, comparable in magnitude to the backbone P peak intensity (Figure \ref{S8}.a-c, pink lines). This indicates that, in addition to electrostatic attraction with phosphate groups, the imidazolium ring accumulates significantly within the minor grooves. The peak position suggests $\pi$-$\pi$ and short-range stacking-type interactions in the minor groove region. Across all concentrations, major groove peaks are weaker than minor groove peaks, confirming that the minor groove is the primary groove-binding site in the dsDNA-rich \textit{ds 90} system. Furthermore, shift of the RDF peaks from $\sim$4 {\AA} to larger distances for the P-[$BMIM$] ring COM and major groove-[$BMIM$] ring COM, arises is due to the unavailability of space filled by IL cations bound within the minor groove.

In \textit{ds 32} (lower panel), the site-specific RDF peak intensity of the imidazolium ring COM and minor groove atoms (pink lines) decreases at higher IL concentrations relative to the backbone P RDF peak. At \textit{100 mM}, some preferential minor groove accumulation is observed (Figure \ref{S8}.d); however, at \textit{250} and \textit{750 mM}, minor groove-[$BMIM$] ring COM RDF peak intensity decreases (Figure \ref{S8}.e-f), and the backbone P-[$BMIM$] ring COM RDF peak dominates (Figure \ref{S8}.d-f), which indicates that for \textit{ds 32} systems electrostatic interactions with phosphate groups become the principal binding mode at higher IL concentrations.

\begin{figure}[H]
    \centering
\includegraphics[width={1\textwidth}]{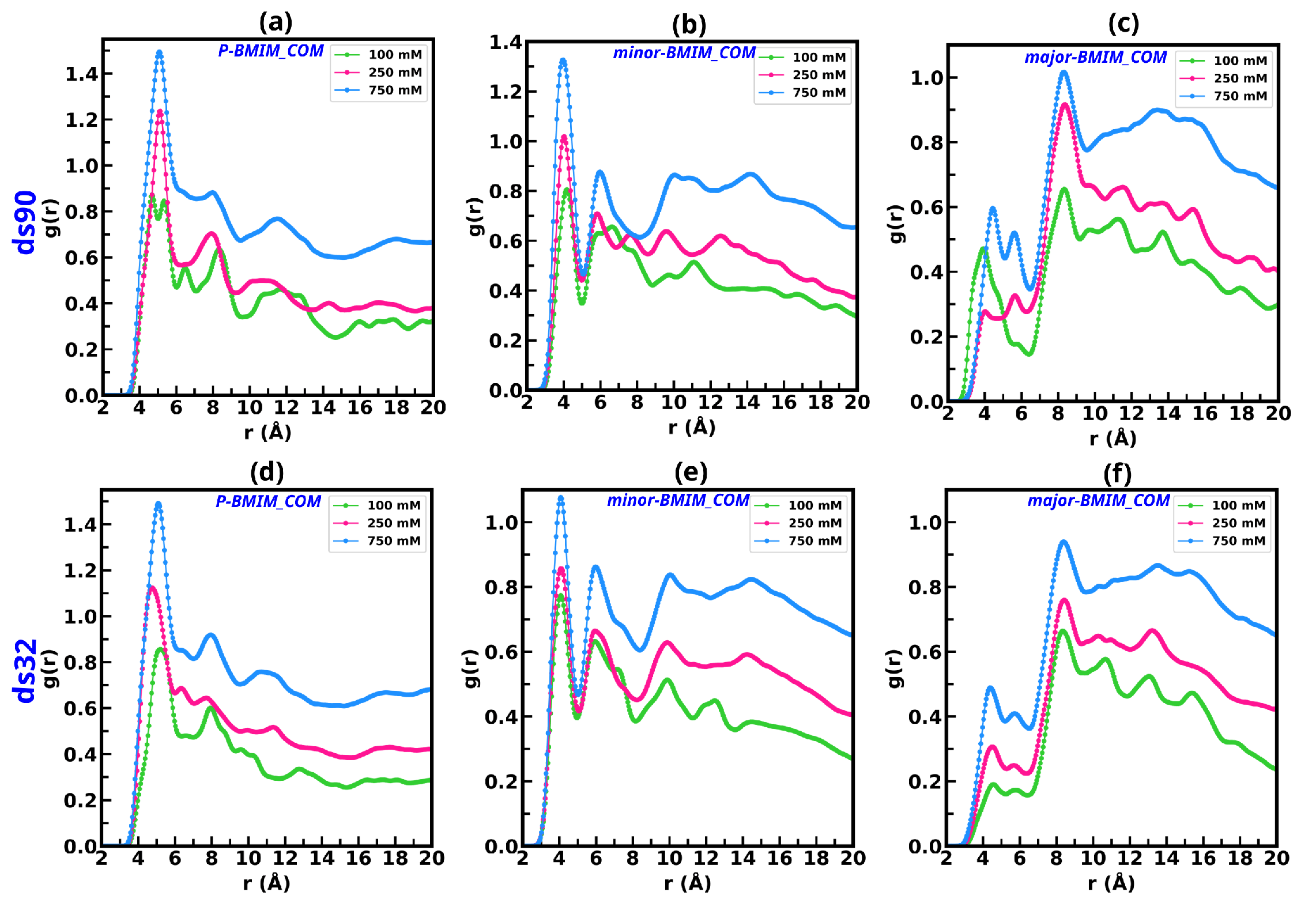}
    \caption{ \textit{Concentration-dependent site-specific radial distribution functions of the [BMIM] imidazolium ring COM around different DNA regions. For dsDNA-rich \textit{ds 90} systems (upper panel): (a) backbone P atoms, (b) minor groove, and (c) major groove. For ssDNA-rich \textit{ds 32} systems (lower panel): (d) backbone, (e) minor groove, and (f) major groove. IL concentrations are 100 mM (green line), 250 mM (pink line), and 750 mM (blue line). All the RDFs are scaled by the IL number density.}}
    \label{S9}
\end{figure}

The concentration-dependent RDF profiles for imidazolium ring COM also show monotonic increase in peak intensity with increasing IL concentration for P backbone (left panel), minor (middle panel), and major groove (right panel) atoms as shown in Figure \ref{S9}.a-c for \textit{ds 32} and Figure \ref{S9}.d-f for \textit{ds 90} systems. 

For \textit{ds 90}, the minor groove-[$BMIM$] ring COM peak increases by $\sim$1.8-fold from 100 to 750 mM (Figure \ref{S9}.b), while the backbone P-[$BMIM$] ring COM peak also increases (Figure \ref{S9}.a). The backbone P-associated IL remains localized at $\sim$5 {\AA}, whereas major groove contributions extend from $\sim$4 to 10 {\AA}, indicating sustained major groove occupancy at higher concentrations (Figure \ref{S9}.c). For ssDNA-rich \textit{ds 32} system, the minor groove-[$BMIM$] ring COM peak increases modestly ($\sim$1.3-fold), whereas backbone P-[$BMIM$] ring COM interactions show a stronger relative enhancement (Figure \ref{S9}.d-e). Major groove contributions remain weak (Figure \ref{S9}.f). 

Overall, these results demonstrate that in dsDNA-rich \textit{ds 90}, groove-mediated interactions, particularly within the minor groove, play a dominant role alongside electrostatic interactions, whereas in ssDNA-rich \textit{ds 32}, IL binding is primarily governed by electrostatic association with the phosphate backbone. 

\section{Electrostatic Potential Maps }
The electrostatic potential, $\phi(\mathbf{r})$, was computed by solving the Poisson equation:

\begin{equation}
\nabla^2 \phi(\mathbf{r}) = -4\pi \sum_i \rho_i(\mathbf{r})
\end{equation}
where the summation runs over all charged atoms, and the charge density $\rho_i(\mathbf{r})$ for each atom is approximated using a spherical Gaussian distribution:

\begin{equation}
\rho_i(\mathbf{r}) = \frac{q_i \omega}{\sqrt{\pi}} \exp\left(-\omega^2 |\mathbf{r} - \mathbf{r}_i|^2\right)
\end{equation}

Here, $q_i$ is the partial charge of the $i$-th atom, $\mathbf{r}_i$ is its position, and $\omega$ is the inverse width of the Gaussian. The instantaneous electrostatic potential $\phi(\mathbf{r})$ was averaged over the final 20 ns of the MD simulation, accounting for all charged atoms in the system.

\begin{figure}[H]
    \centering
\includegraphics[width={1\textwidth}]{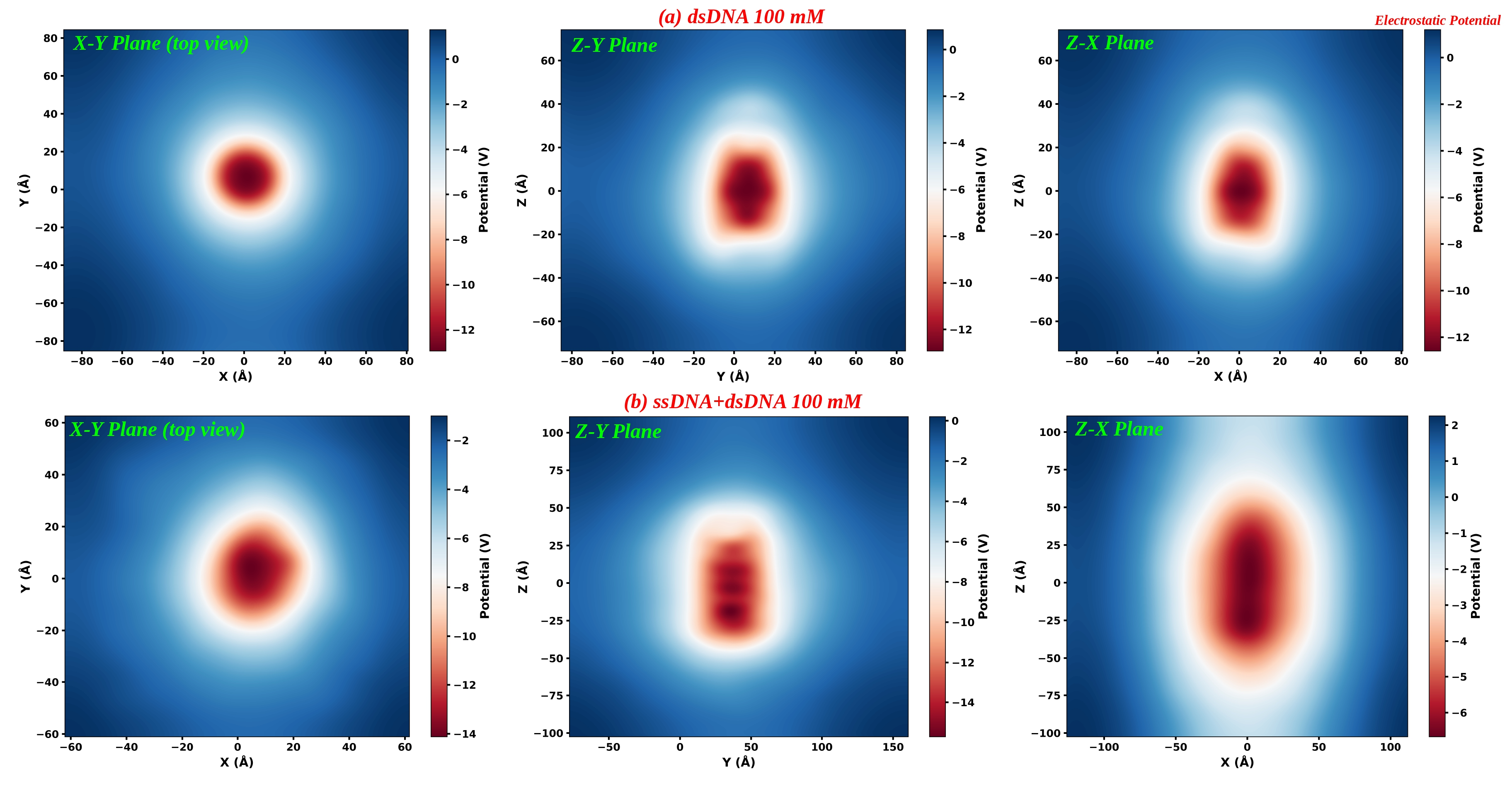}
    \caption{ \textit{ Representation of electrostatic potential distribution map: (a) For dsDNA with 100 mM [BMIM]acetate, 2D maps (x–y, z–y, z–x planes) show deep brown color regions indicating strong negative potentials from dsDNA. BMIM cations partially neutralize these via groove binding and electrostatic interactions. (b) In ssDNA-rich systems (DNA+BMIM+Acetate), potential maps reveal high-density regions in deep brown color. The cationic BMIMs primarily interact with DNA backbones via electrostatic interactions, expanding neutral zones (white regions) and reducing groove binding relative to dsDNA-rich systems. Electrostatic interactions bring the potential closer to zero by screening the negative potential of DNA backbones in the presence of cationic BMIMs. The color scale bar on the right side indicates electrostatic potential in volts (V). The first column represents the top view in the X-Y plane, and the second and third columns represent the side views in the Z-Y and Z-X planes of the systems in the figure. Overall, electrostatics is more prominent in ssDNA-rich systems.}}
    \label{S10}
\end{figure}

\begin{figure}[H]
    \centering
\includegraphics[width={1\textwidth}]{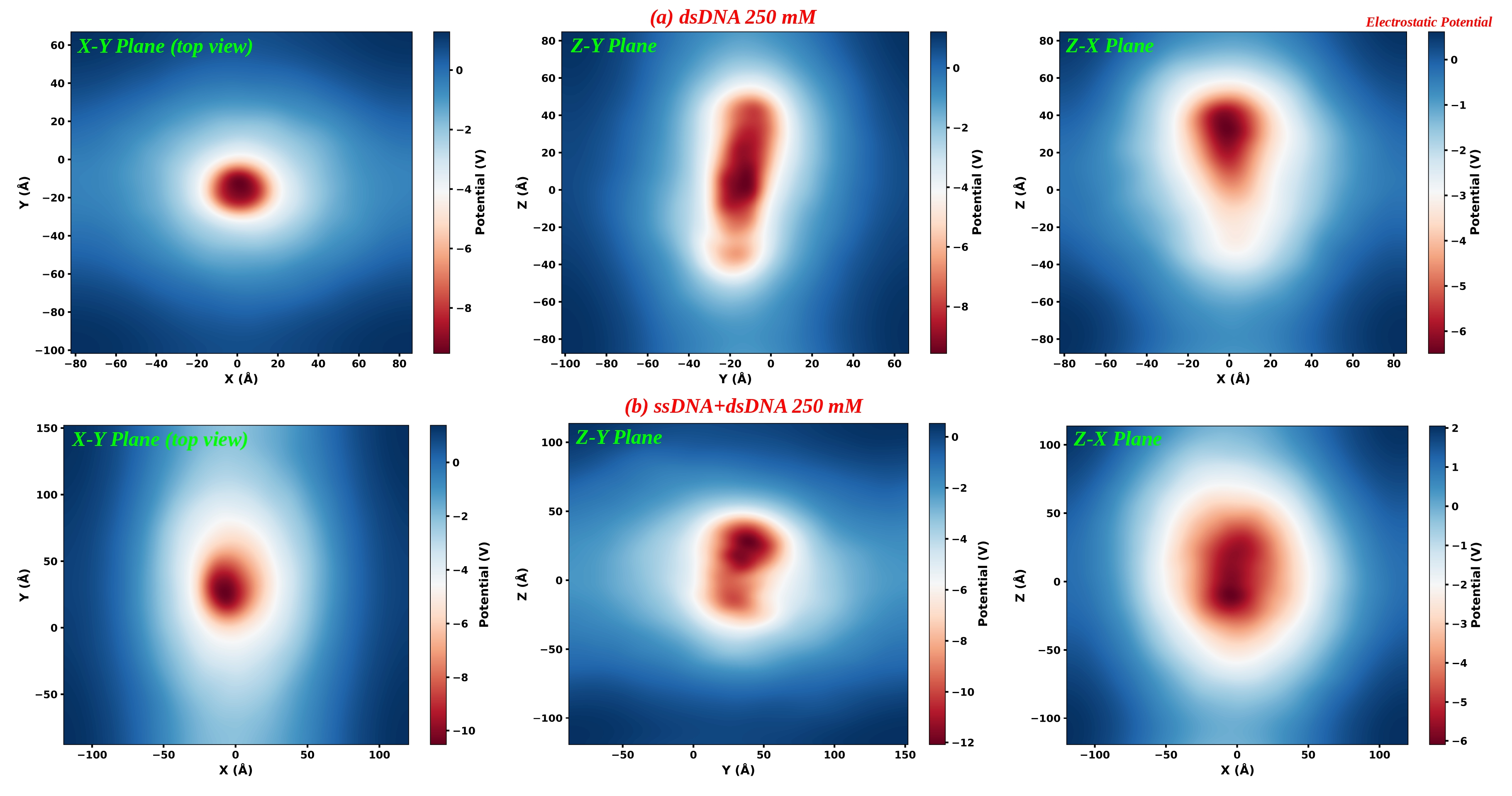}
    \caption{ \textit{Electrostatic potential maps- (a) For dsDNA-rich systems at 250 mM [BMIM]acetate, electrostatic potential maps reveal groove binding, with reduced spreading of deep brown regions across all planes. The X-Y plane shows the top view, while the Z-Y and Z-X planes present side views of the potential distribution. (b) In ssDNA-rich systems, stronger backbone binding of cationic ILs leads to greater neutralization of negative potential, as evidenced by the increased white regions indicating enhanced electrostatic screening. The color scale represents potential in volts (V).}}
    \label{S11}
\end{figure}

\begin{figure}[H]
    \centering
\includegraphics[width={1\textwidth}]{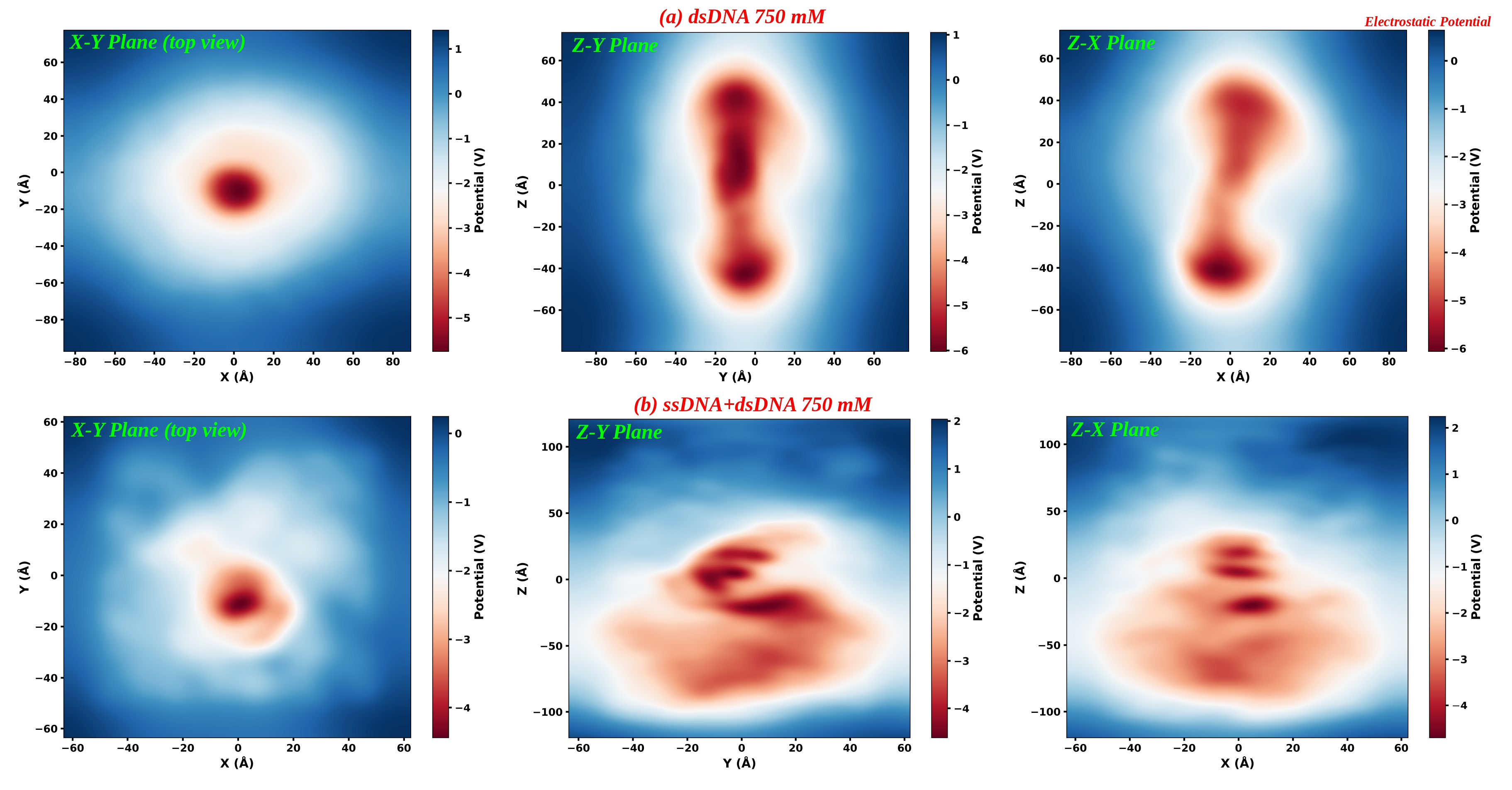}
    \caption{ \textit{Electrostatic potential maps- (a) For dsDNA-rich systems at 750 mM [BMIM]acetate, 2D maps of total electrostatic potential from all charged atoms reveal prominent groove binding, as evident in the upper panel across all three planes. (b) In contrast, ssDNA-rich systems (lower panel) exhibit dominant electrostatic interactions between DNA and cationic BMIM. This leads to chain collapse and significant neutralization of the negative potential, resulting in widespread white colored regions across X-Y, Z-Y, and Z-X planes. The increased electrostatic screening reduces groove bindings in ssDNA-rich systems, unlike the well-defined groove bindings observed in dsDNA-rich systems. The color scale bar on the right side gives the electrostatic potential in units of volts (V).}}
    \label{S12}
\end{figure}


\end{document}